*Multi-messenger nano-probes of hidden magnetism in a strained manganite*

A. S. McLeod [1,2,*,†], J. Zhang [1,*], M. Q. Gu [3], F. Jin [4], G. Zhang [1], K. W. Post [1], X. G. Zhao[5],

A. J. Millis [2], W. Wu [4,6], J. M. Rondinelli [3], R. D. Averitt [1], & D. N. Basov [1,2]

[1] Department of Physics, University of California San Diego, La Jolla, California 92093, USA

[2] Department of Physics, Columbia University 538 West 120th Street, New York, New York 10027, USA

[3] Department of Materials Science and Engineering, Northwestern University, Evanston, Illinois 60208, USA

[4] Hefei National Laboratory for Physical Sciences at the Microscale, University of Science and Technology of China, Hefei 230026, China

[5] Department of Mechanical Engineering, Boston University, Boston, Massachusetts 02215, USA

[6] Institute of Physical Science and Information Technology, Anhui University, Hefei 230026, China

**Abstract:**

The ground state properties of correlated electron systems can be extraordinarily sensitive to external stimuli, such as temperature, strain, and electromagnetic fields, offering abundant platforms for functional materials. We present a metastable and reversible photoinduced ferromagnetic transition in strained films of the doped manganite $La_{2/3}Ca_{1/3}MnO_3$ (LCMO). Using the novel multi-messenger combination of atomic force microscopy, cryogenic scanning near-field optical microscopy, magnetic force microscopy, and ultrafast laser excitation, we demonstrate both "writing" and "erasing" of a metastable ferromagnetic metal phase with nanometer-resolved finesse. By tracking both optical conductivity and magnetism at the nano-scale, we reveal how spontaneous strain underlies the thermal stability, persistence, and reversal of this photoinduced metal. Our first-principles electronic structure calculations reveal how an epitaxially engineered Jahn-

[*] These authors contributed equally to the present work.

Corresponding author:
[†] am4734@columbia.edu

Teller distortion can stabilize nearly degenerate antiferromagnetic insulator and ferromagnetic metal phases. We propose a Ginzburg-Landau description to rationalize the co-active interplay of strain, lattice distortion, and magnetism we resolve in strained LCMO, thus guiding future functional engineering of epitaxial oxides like manganites into the regime of phase-programmable materials.

Ultrafast all-optical control of the insulator-metal transition in correlated electron systems[1] remains a coveted route towards reconfigurable functional materials.[2] Although transient photoinduced phase transitions can expose interactions among competing orders in complex systems,[3] persistent all-optical and reversible switching is desirable for on-demand device functionalities.[4] For example, the colossal magnetoresistive (CMR) manganites ($AE_{1-x}RE_xMnO_3$, AE: alkali earth, RE: rare earth) show magnetic and electronic properties that depend strongly on the microscopic geometrical configuration of the comprising $MnO_6$ octahedral network.[5–8] This pliancy affords fertile ground for photoinduced manipulation of the equilibrium state.[9,10] In particular, the accompaniment of magnetism by spontaneous strain in Ca-doped $La_{1-x}Ca_xMnO_3$[11,12] raises the possibility for interplay between strain-coupling (often termed "coelasticity"[13]) in this compound and the photoinduced insulator-metal transition.[14] Here we study a 26 nm thick $La_{2/3}Ca_{1/3}MnO_3$ (LCMO) thin film grown on $NdGaO_3$ (001) substrate.[15] Although first exhibiting a low-temperature insulator to metal transition (IMT) common to the bulk system at $T$~250K, annealing these films in an oxygen environment coherently accommodates their epitaxy to the $NdGaO_3$ substrate, stabilizing a persistent antiferromagnetic insulating (AFI) phase.[15] Strikingly, these insulating LCMO films show extreme susceptibility to ultra-short pulsed laser excitation (<200 fs), driving a persistent phase transition into a metallic state.[14] Although the IMT in rare-earth manganites is commonly accompanied by ferromagnetic (FM) order, persistent photoinduced magnetism in this material awaits direct experimental confirmation. Moreover, the reversibility and stability of this photoinduced phase remain unexplored at the nanometer length scales characteristic of phase separation in manganites[16–21] – a phenomenon underlying colossal magneto-resistance (CMR). In this work we demonstrate how inherent strain-coupling of the photoinduced IMT in a strained

Corresponding author:
† am4734@columbia.edu

manganite explains the cooperative stability of the transition, its tendency towards phase coexistence, and its reversibility through spatially selective delivery of nano-scale strain.

## Multi-messenger nano-probes & photo-excitation of hidden ferromagnetic metal

To optically monitor the photoinduced IMT at the nano-scale, we leverage cryogenic scattering-type scanning near-field optical microscopy (cryo-SNOM). In this atomic force microscope (AFM)-based imaging technique, the optical signal $S$ registered by SNOM affords a 10 nm-resolved probe of low-energy metallic ("Drude") conductivity (see Methods). Meanwhile, magnetic force microscopy (MFM) provides a nano-resolved probe of local magnetic moments in ferro- and ferrimagnets, routinely applied to visualize magnetic phase separation in manganites[16,22,23]. Using a magnetic near-field probe, we combine both SNOM and MFM techniques for collocated multi-messenger imaging of nano-scale magnetic and optical properties across the IMT.

We first elucidate the nature of the "hidden" metallic state in LCMO by presenting correlative SNOM and MFM measurements of the strained film at temperatures $T$<240K where the bulk system conventionally displays a ferromagnetic metal ground state. Fig. 1a presents the AFM topography at $T$=96K of a film region characterized by a pair of crossed cracks shown as a dark "trough" in surface topography and a faint step highlighted by the dashed white line; these presumably formed during film growth. Fig. 1b presents the corresponding SNOM image: bright yellow signifies a SNOM signal $S$ indicating a strong metallic response, whereas dark blue reflects the insulating response of the AFI ground state. It can be immediately seen that the locations of cracks locally "restore" a metallic phase with ferromagnetism identifiable by MFM (Fig. 1c). Here the dark/light profile of magnetic forces registered across each crack associated with "fringe" magnetic fields indicating an in-plane ferromagnetic moment. Although the "hidden" ferromagnetic metal (FMM) remains localized to these cracks, its magnetic moment and metallic conductivity are found to survive up to T=200K, vanishing monotonically with increasing temperature (see Supplementary Information or "SI" for additional images). Moreover, Figs. 1b&c also resolve faintly metallic FM domains ~100 nm in size sparsely scattered throughout the film; their striped texture aligned perpendicular to the LCMO $b$-axis resembles the magnetic domains resolved in

Corresponding author:
† am4734@columbia.edu

related strained films.[23] This largely insulating behavior contrasts with the insulator-metal coexistence we observe in partially annealed LCMO films below T=100K, as well as the full metallicity detected in incoherently strained films[15,24] (nano-imaging of additional films in SI).

Having explored the equilibrium state of our strained film, we now proceed to examine the photoinduced insulator-metal transition with nanoscale resolution. To achieve this, a controlled number of ultrafast (150 fs) pulses with tunable fluence at 1.5 eV from an amplified Ti:Sapphire laser are directed into the cryo-SNOM chamber, where they are focused by the SNOM optics to the sample surface directly adjacent to our SNOM probe (Methods), as shown schematically in Fig. 1d. After illuminating the sample, we conduct local SNOM and MFM imaging to resolve the spatial extent of the photoinduced phase transition. Fig. 1e presents SNOM images of local metallicity induced by a sequence of excitation pulses at powers of 5, 15, and 30 μW, corresponding to fluences of a few mJ $cm^{-2}$ at the focused excitation spot, delivered to the sample at $T$=70K at locations marked by dashed ovals. Here a progression of metallicity results from increased fluence, correlated with the simultaneously acquired MFM images in Fig. 1f. First, we find a threshold excitation fluence of 5 μW (~1 mJ $cm^{-2}$) is necessary to achieve any photoinduced change, resulting only in formation of sparse disconnected stripe-like domains, with a weak and disordered magnetic response (Fig. 1e, left-most panel), similar to those isolated domains (false-colored green) resolved in Fig. 1b. Meanwhile, 15 μW (~4 mJ $cm^{-2}$) triggers a larger (~10 microns) single FMM domain, whose weakly metallic edges remain magnetically disordered. Fluence of 30 μW excites a coherent metallic domain whose shape repeats the spatial profile of the excitation field (Fig. 1e, right-most panel). Importantly, even this uniformly magnetized FM domain presents an outer boundary "frayed" in an incoherent stripe-like pattern (e.g. Fig. 1e&f, center panels) reminiscent of that among incipient metallic stripes (cf. Fig. 1b&c); we later revisit the origins for this spatial texture.

Images in Fig. 1 reveal the photoinduced state achieved through excitation with a train of approximately one thousand laser pulses (~1s exposure). Meanwhile, to unveil pulse-to-pulse progression of the photoinduced IMT, we also examine the hidden FMM under incremental stimulation from discrete laser pulses. Fig. 1g presents metallicity of

Corresponding author:
† am4734@columbia.edu

photoinduced domains produced by an increasing number of sequential laser pulses at 20 μW (top row of panel, number of pulses indicated), complemented by simultaneously acquired magnetization maps in Fig. 1h. The optical pulses drive a cumulative growth in both domain size as well as overall metallicity, as plotted in Fig. 1i&j, respectively. Meanwhile, higher fluence (40 μW, ~8 mJ $cm^{-2}$; bottom rows of Figs. 1g&h) induces much more rapid growth of metallic domains to a size double that achieved at 20 μW. Combined with our nano-imaging of FMM re-emerging at cracks, these measurements display four aspects of the hidden photo-excited magnetic metal: i) application or release of epitaxial strain controls stability of the FMM state at the nano-scale; ii) the maximum size of a photoinduced FMM domains is set by the sample area exposed to excitation optical fields exceeding a threshold amplitude $E_{\mathrm{thresh}}$; iii) growth in both size and metallicity of domains is self-assisted, in that incremental laser pulses achieve a cumulative switching efficiency; and iv) excitation at sufficiently high fluence (e.g. 40 μW) impairs the metallicity of domains, as shown in Fig. 1i&j for > 10 pulses. As we will show, these phenomena can be understood within a thermodynamic free energy model for the strained LCMO film, revealing that strain plays a key role both for the surprising metastability and photoexcitation susceptibility of the hidden ferromagnetic metal.

## *Ab initio* and Landau description of the strain-coupled phase transition

To clarify interplay between epitaxial strain and the hidden FMM revealed by our nano-imaging, we performed density functional theory calculations to predict the properties[5] of bi-axially strained LCMO at an *ab initio* level, accounting for electron correlations through the on-site Coulomb interaction (DFT+U)[25]. Bi-axial strain is expected to impart enhanced $Q_2$ Jahn-Teller distortion to $MnO_6$ octahedra of the LCMO lattice (Fig. 2c)[26], thus reducing the electronic bandwidth and enhancing the influence of electronic correlations with potential to open a band-gap.[21,27–29] Accordingly, Fig. 2d presents electronic calculations of the strained lattice revealing that the antiferromagnetic phase attains lower energy than the ferromagnetic phase when tensile *b*-axis strain exceeds a critical threshold. Although larger than the experimental strain condition, the calculated strain threshold (8%) is within the same order of magnitude. At such critical strains, these

Corresponding author:
† am4734@columbia.edu

competing magnetic phases are predicted to stabilize at distinct lattice configurations with respect to the $Q_2$ distortion. Spin-resolved densities of states computed for these strained phases are shown in the Fig 2d inset, revealing a >100 meV charge gap in the strain-stabilized AFI phase. Although these variations in $Q_2$ distortion remain to be experimentally confirmed, application or relief of strain towards their thresholds of stability is expected to enable insulator-metal transitions between the AFI and FMM phases dressed by a first-order lattice transition (see SI for further characteristics of these phases computed at distinct values of b-axis strain). Our *ab initio* calculations can thus contextualize the observed emergence of "hidden" nano-scale FMM at cracks in the film (Fig. 2a) by considering the local relief of epitaxial strain. Fig. 2b presents a finite element simulation (Methods) of the strain distribution local to a crack in our film, showing that tensile epitaxial strain along the manganite *b*-axis is locally relieved at the location of a crack (dashed line) within proximity of a few film thicknesses (~50 nm), highlighted by the shaded line plot above the schematic LCMO film. This simulation accords with our SNOM and MFM measurements that resolve the hidden ferromagnetic metal phase within about 100 nm of these cracks, establishing a direct connection between the local nano-scale strain environment and stability of the AFI ground state. Moreover, we speculate that the "incipient" metallic domains (false-colored green) appearing before any photo-excitation in Figs. 1b&c result from incomplete relief of epitaxial strain at e.g. crystallographic defects, since their appearance is spatially reproducible. Combined with this simulation, our *ab initio* results imply that epitaxial strain suppresses ferromagnetism by promoting $Q_2$ Jahn-Teller distortion to strain-stabilize a AFI ground state.

These findings motivate us to formulate a Ginzburg-Landau (GL) theory explaining the relative stability and photo-susceptibility of magnetic phases in our strained manganite film. Landau theories previously applied to rare earth manganites have considered the amplitude and phase of charge order of the insulating state together with the ferromagnetic moment $M$ of the metallic state as relevant degrees of freedom[14,30] Although key to the novel properties of strained LCMO, their interplay with strain remains underexplored, motivating our development of a modified GL theory with two components. First, by associating the AFI phase with the distorted structure reflected in the ab initio results presented here (Fig. 2c&d), we associate the amplitude $Q$ of a multi-modal lattice distortion (dominated by

Corresponding author:
† am4734@columbia.edu

increase in the $Q_2$ Jahn-Teller octahedral distortion) with the order parameter of the charge-ordered insulator, as in Ref. [14]. The free energy density of the LCMO electronic subsystem can be thereby expressed in terms of scaled order parameters $Q$ and $M$, whose magneto-elastic coupling produces an energy barrier between metastable FMM ($M \neq 0, Q = 0$) and AFI ($M = 0, Q \neq 0$) phases (see Methods and SI). Since our *ab initio* calculations imply that a AFI phase can be stabilized through *b*-axis tensile strain and associated increase in $Q_2$ Jahn-Teller distortion (Fig. 2c), we postulate a contribution $F_\varepsilon$ to the overall free energy density arising from symmetry-allowed coupling between the Jahn-Teller distortion and strain, which to lowest order in $Q$ is (Methods):

$$F_\varepsilon \approx -Q^2\, \delta\varepsilon : K : (\varepsilon + \varepsilon_{\mathrm{ep}}). \qquad \text{(Eq.1)}$$

Here $K$ denotes an adimensionalized elastic stiffness of the film and substrate, whereas $\varepsilon$ denotes the local strain tensor relative to the equilibrium elastic configuration of the substrate (imparting the epitaxial constraint $\langle\varepsilon\rangle$=0 of net zero areal strain) and $\varepsilon_{\mathrm{ep}}$ quantifies the epitaxial strain on the film (see SI). The essential term in Eq. 2 is $\delta\varepsilon$, a "spontaneous strain"[31] denoting the relative difference in relaxed (unclamped) LCMO lattice constants between AFI and FM phases at low temperature. This spontaneous strain was found to be significant (~1%) for over-doped LCMO upon entry to the insulating state, growing even larger for $T<T_M$, with $T_M$ the Curie temperature.[12] At this level of theory, application of epitaxial strain $\varepsilon_{\mathrm{ep}}$ can extrinsically boost the transition temperature $T_Q$ into the AFI ($Q^2 > 0$) phase state by an amount $\delta T_Q/T_Q \propto \delta\sigma : \varepsilon_{\mathrm{ep}}$, where $\delta\sigma \equiv K : \delta\varepsilon$ is the "spontaneous stress" on the lattice associated with $\delta\varepsilon$. This mechanism affords a pathway towards a metastable AFI phase provided that $T_Q$ is pushed to exceed $T_M$.

We now address the photoinduced IMT in the context of our GL theory. Fig. 3a depicts the consequent free energy density applicable to our coherently strained LCMO film. Minima in this surface correspond with metastable AFI ($Q\neq 0$) and FMM ($M\neq 0$) phases, of which the former becomes the strain-mediated and thermally accessible ground state. The difference in Mn-$e_g$ orbital occupation between AFI and FMM phases obtained by our *ab initio* calculations (Fig. 2d & SI) suggests an optically-triggered phase transition resulting from collective redistribution of electrons within a unique orbital manifold, as proposed previously.[14] Here ultrafast laser excitation tuned to the inter-site Mn-$e_g$

Corresponding author:
† am4734@columbia.edu

transition is proposed to delocalize charge ordered carriers, triggering a momentary relaxation of the $Q^2 > 0$ state. As suggested by Fig. 3a, the efficiency of photo-excitation should depend sensitively on the free energy barrier between FMM ($Q$=$0$) and AFI ($Q$≠$0$) points of stability, reflected in the critical fluence threshold evidenced by Fig. 1e&f for the photoinduced IMT. As reflected in our *ab initio* results, coupling to the local strain tensor $\varepsilon$ (Eq. 1) at fixed $T$ necessarily dresses the IMT with an anomalously abrupt change in $MnO_6$ octahedral configuration, signature of a discontinuous strain-coupled transition[13]. This coupling predicts dramatic ramifications for the growth in both size and metallicity of photoinduced domains under incremental laser pulses resolved in Figs. 1g&h. Fig. 3b presents the result of a phase-field modeling simulation of a metastable metallic domain minimizing the strain-coupled free energy density (Eqs. 2&3 in Methods; see SI for simulation details). The top panel depicts metallic regions ($Q$=0) in yellow and insulating regions ($Q$=$Q_0$) in blue, whereas the bottom panel presents the self-consistently computed strain field associated with spontaneous strain of the AFI phase. Here data are presented as the difference in local elastic energy cost $\Delta F_\varepsilon = F_\varepsilon^{\text{ins}} - F_\varepsilon^{\text{met}}$ between insulating and metallic states relative to the energy scale $\delta E_\varepsilon = \delta\sigma : \delta\varepsilon$ of accommodation strain (see Methods and SI). Fig. 3b implies that accommodation strain resulting from the spontaneous stress $\delta\sigma$ around a FMM domain can locally i) counter-act the epitaxial strain, ii) raise elastic energy associated with the AFI phase, and iii) dynamically reduce the energy barrier for a photoinduced phase transition. Fig. 3c presents the total free energy density at a sequence of local strain environments approaching the metallic domain from the outside (blue to red), showing a "softening" of AFI stability reminiscent of the strain relief at cracks discussed earlier. Strain-coupling of the IMT can thus explain growth in size and metallicity of sequentially photo-excited domains resolved here at the nano-scale: domain formation co-actively reduces elastic barriers to subsequent photo-excitation of the FMM phase through localized relief of epitaxial strain energy near insulator-metal domain boundaries (blue in Fig. 3c). Assuming merely that photo-excitation above a critical field $E_{\text{crit}}$ can overcome the initial energy barrier between FMM and AFI phases, Fig. 3d presents a simple simulation of self-consistent FMM domain growth under incremental photo-excitation at $E_{\text{crit}}$ and $2E_{\text{crit}}$ subject to the realistic accommodation strain field of Fig. 3c, with $A_{\text{pulse}}$ the FWHM area of the photo-excitation spot (see SI for details of the model).

Corresponding author:
† am4734@columbia.edu

Alternative simulations assuming e.g. a long-range magnetic dipole field originating from the volume of the FMM domain (coupling to $M$ at an energy scale comparable to $\Delta F_{\varepsilon}$) provides far worse comparisons to the experimental data (Fig. 1i). In the context of GL theory, strain-coupling (Eq. 1) of the IMT and attendant accommodation strain thus prove fundamental to both persistence and co-active growth of photo-excited magnetic domains in strained LCMO.

**Thermal melting of the photoinduced ferromagnetic metal**

The reduction of metallicity resolved among photoinduced domains excited at sufficiently high fluence (e.g. 40 uW, see Fig. 1g&h) hints at thermal "melting" of the hidden FM order. Here we propose that photo-excitation raises the local temperature towards instability of the FMM. To directly resolve the melting transition, we performed nano-imaging of a large (100 x 100 microns) domain "written" at $T$=70K with an optimal fluence of 30 uW over increasing temperatures. Fig. 4a&b present thermal evolution of metallicity and magnetization for the photoinduced phase within a large (40 x 40 microns) fixed field of view close to the edge of the written domain. Metallicity first weakens at $T$=92K, giving way at 94K to abundant electronic phase coexistence, a hallmark feature of colossal magnetoresistance.[20,21] The striped domain morphology of melting FMM resembles the broken metallicity observed within metallic domains (Fig. 1g) photoinduced at high fluence ($\approx$8 mJ cm$^{-2}$). These domains were evidently photo-thermalized into an inhomogeneous state above $T$=92K where AFI and FM phases coexist. Moreover, the striped domain texture highlighted in Fig. 4c resembles that observed in Figs. 1e&f, suggesting an underlying physical origin with similarity to textures uncovered amidst IMTs of other transition metal oxides, such as $V_2O_3$[32] and $VO_2$[33,34] and otherwise resolved as microscopic "tweed" patterns among twinned crystals.[35,36] In the SI, we detail how these hallmark patterns further signify how strain-coupling underlies this unconventional IMT in strained LCMO. In particular, oriented metallic stripes perpendicular to the LCMO b-axis reflect anisotropy in the spontaneous strain $\delta\varepsilon$. Our GL theory predicts that their characteristic periodicity $\xi$ (Fig. 4c) emerges spontaneously to minimize accommodation strain of coexisting FMM and AFI phases within the film balanced against the energetic cost of FM / AFI domain boundaries, as detailed with additional simulations presented in the SI.

Corresponding author:
† am4734@columbia.edu

The correlative SNOM and MFM images presented in Fig. 4 afford an unprecedented opportunity to locally track magnetism and metallicity of the "hidden" FMM during its reentrant metal-insulator transition. Fig. 4d resolves the overall phase transition through a histogram representation of SNOM signals $S$ (viz. metallicity)[32], relative to that of the insulating state, presented at increasing temperatures. Although these histograms reflect a bi-modal component clearly distinguishing insulator from metal, they also implicate a continuous drop in metallicity upon thermal reentry to the AFI phase. While this phenomenology superficially resembles a second-order transition to the paramagnetic insulator phase at $T_{\mathrm{M}}$ (described by Eq. 2 in Methods), the observed transition at $T_M \approx 110\ K$ implies an immense renormalization of the Curie temperature down from the bulk value of 250 K – much lower even than the transition temperature associated with FMM at strain-relieved cracks (Fig. 1a-c). We attribute this reduction to Jahn-Teller distortion under coherent biaxial strain[37], a magnetostriction exceeding that reported for other manganite films[38,39]. Fig. 4e provides more detail by identifying classes of pixels behaving similarly during the thermal transition by virtue of their characteristic "transition temperature" $T_C$, defined as that where metallicity measured by SNOM reduces to half the full FMM value. We present the thermal evolution of different regions of the film in a phase space of metallicity ($S$ measured by SNOM) and magnetic moment ($M$ measured by MFM). Traces presented in Fig. 4e (smooth curves are guides to the eye) are color-coded to characteristic values of $T_c$ ranging from ~90K to ~100K, between which the scaling of magnetic moment to metallicity is found to vary widely. Although overall thermal evolution of the magnetization and conductivity follows trends seemingly predicted by the double-exchange paradigm,[40] our nano-resolved measurements reveal widely heterogeneous character of this reentrant metal-insulator transition. Whereas FMM regions persisting to high $T_c$ (~97K) abide a continuous commensurate decrease in both metallicity and magnetization, those with lower $T_c$ (~89K) present a rapid drop in optical conductivity preceding the magnetic transition. We propose that spatial variations in electron-phonon coupling,[41,42] perhaps associated with the inhomogeneous texture of accommodation strain, might account for these variations. This hypothesis motivates future explorations of magnetism in this photoinduced metastable metal.

**Nano-scale erasure of the photoinduced ferromagnetic metal**

Corresponding author:
† am4734@columbia.edu

The possibility to selectively trigger an abrupt transformation through external stress comprises the most functional hallmark of strain-coupled ferro- and coelastic transitions.[43,44] Here we demonstrate this feature for the IMT among photoinduced FMM domains, thus establishing complete functional reconfigurability of the electronic and magnetic state of this strained manganite with nano-scale finesse. Since distinct $MnO_6$ octahedral configurations are predicted for the strained FMM and AFI states explored here (Fig. 2c) and manipulable by $b$-axis strain, we examined the effect of compressive $c$-axis stress $\sigma_{cc}^{tip}$ applied from the tip of our cantilevered AFM probe to the photoinduced FMM phase. By the sizable Poisson effect reported in thin film manganites, we expect our applied stress to impart further tensile strain into the $ab$-plane with $\approx 40\,\%$ efficiency,[45] with potential to destabilize the metastable FMM especially at temperatures poised within 20K of the melting temperature $T_{\mathrm{M}}$ (Fig. 4a&b). Fig. 5a schematically presents SNOM imaging of a photoinduced domain at such temperatures ($T$=90K) before stress application. Meanwhile, Fig. 5b displays the outcome of "writing" stress – by raster-scanning the sample in mechanical contact with the AFM probe – onto a selected 3 micron-square region (dashed area) of FMM with three sequentially applied levels of stress delivered by our probe (stress calibration discussed in Methods). By $\sigma_{cc}^{tip} \approx 125$ MPa the FMM phase suffers a decrease in metallicity, whereas 250 MPa locally induces a large domain of reentrant AFI phase. By $\sigma_{cc}^{tip} \approx 500$ MPa the FMM phase is almost entirely "erased", leaving behind only isolated puddles of remnant metallicity within the region of erasure. Here, peak compressive stress within the film is expected to exceed 2 GPa (SI) and to impart $b$-axis tensile strain comparable to the epitaxial value $\varepsilon_{\mathrm{ep}}$. The GL description (Eq. 1) predicts such externally imposed strain $\varepsilon$ is sufficient to completely destabilize the $Q = 0$ (FMM) phase, triggering an abrupt transition to the equilibrium $Q \neq 0$ AFI state (see SI). Accordingly, we find that $\sigma_{cc}^{tip} \approx 500$ MPa can consistently erase the FMM phase with $< 500$ nm-resolved finesse, as demonstrated in Fig. 5c via three consecutive "erasures" (dashed white regions). Concurrent MFM imaging (Fig. 5d) confirms that ferromagnetism is likewise erased with sub-micron control. Although the AFM probe in our microscope is held at elevated temperature compared with the sample, a purely thermal mechanism of erasure based on local heating

Corresponding author:
† am4734@columbia.edu

fails to account for our full findings, although the local erasure may be thermally assisted (discussed in SI).

## Outlook

The novel reversal pathway established here for the magnetic transition in $La_{2/3}Ca_{1/3}MO_3$ excels in spatial selectivity over previous demonstrations of non-volatile magnetic switching in related compounds,[46,47] inspiring a new paradigm for complete opto-elastic control over magnetism. Consistent with our *ab initio* results and Ginzburg-Landau description, Fig. 5e establishes a heuristic phase diagram summarizing our findings for the insulator-metal transition in strained LCMO. Nano-imaging of additional films associated with strains/temperatures marked by the gray symbols are presented in SI. Our nano-imaging results map how the limits of thermodynamic stability for the hidden ferromagnetic metal are defined by 1) temperature, 2) photoexcitation fluence, and 3) epitaxial/external strain, revealing how co-active interplay of these factors underlies our selective control over both magnetism and metallicity. In particular, strain-coupling and strain-mediated growth of the hidden ferromagnetic metal uncovered here imply extreme pliancy to epitaxial and external stress, owing to strongly lattice-dependent electronic properties among manganites. Our *ab initio* prediction of an insulating phase stabilized at critical strains exceeding the epitaxial strain condition motivates future studies addressing importance of the film-substrate interface and charge ordering. Nevertheless, our Ginzburg-Landau description is generalizable, and should inform future design principles for functional devices that leverage complete, non-volatile, and reversible photo-elastic switching of the IMT in correlated electron materials – particularly in similarly strained manganite films possessing higher "intrinsic" Curie temperature such as $La_{2/3}Sr_{1/3}MnO_3$ or $La_{2/3}Ba_{1/3}MnO_3$.[21] Our results moreover demonstrate an unprecedented multi-messenger nano-probe for interrogating metastable photoinduced phase transitions with nano-scale selectivity, thus mapping intermediary regimes of electronic and magnetic phase coexistence otherwise inscrutable with conventional bulk probes. Our methodology should next enable precise manipulation and spectroscopic interrogation of local electronic properties among hidden phases beyond the strained ferromagnetic metal studied here. Moreover, we envision the immediate extension of cryogenic SNOM for nano-resolved investigation of photoinduced phase

Corresponding author:
† am4734@columbia.edu

transitions in other fundamental correlated electron materials, including Mott insulators,[48] transition metal chalcogenides,[49] and superconductors.[50]

Corresponding author:
† am4734@columbia.edu

## Methods

### Cryogenic scanning near-field optical microscopy (cryo-SNOM)

Scanning near-field optical microscopy (SNOM) is an atomic force microscope (AFM)-based technique enables imaging of surface optical properties at variable temperatures[32,51,52] below the diffraction limit[53], with a resolution strictly limited only by the geometric sharpness of the metallic AFM probe. We present 25 nm-resolved imaging of the locally back-scattered near-field signal amplitude (abbreviated to *SNOM signal*, or *S*) collected at low temperatures (down to $T$=50K) using a custom-designed cryogenic near-field optical microscope (cryo-SNOM).[54] In these measurements, focused infrared light is incident upon and scattered from the metallic tip of an atomic force microscope (AFM) probe (240AC-MA, Mikromasch USA) oscillating at a frequency ~70 kHz near the sample surface while the microscope is operated in amplitude modulation AFM feedback. The back-scattered radiation from the probe encodes information about the optical permittivity of the sample at the frequency of the laser source. Back-scattered radiation is registered by a liquid nitrogen-cooled mercury cadmium telluride photodetector and resolved from the background through Michelson interferometry in a pseudo-heterodyne detection scheme. To a first approximation, the amplitude of back-scattered radiation modulated at high harmonics ($n \geq 2$) of the cantilever oscillation frequency provides a measure of the local near-field optical response of the sample and, by implication, its optical conductivity resolved at the 20-nm scale[15,47]. The second harmonic signal at $n$=2 is presented throughout this work. Since the bandgap of our LCMO film in the AFI state is about 1 eV,[14] nano-imaging with infrared (IR) light at 100 meV provides an unambiguous local probe of manganite metallicity by way of its infrared free-carrier ("Drude") response.

### Magnetic force microscopy

Magnetic force microscopy (MFM) resolves relative shifts ($\Delta f$, on order of Hz) in the resonance frequency of a cantilevered AFM probe coated in a ferromagnetic metal to infer the presence and strength of ferromagnetic moments in the sample.[55] Localized magnetic dipole interactions between the probe tip and the sample surface produce positive frequency shifts when repulsive, indicating counter-oriented magnetic fields at the sample

Corresponding author:
† am4734@columbia.edu

surface relative to the probe's magnetic moment, whereas negative shifts reflect an attractive interaction and co-oriented fields. MFM measurements in the present work utilized a magnetic probe (240AC-MA, Mikromasch USA) with a magnetic Co-alloy coating characterized by a moment of $\approx 10^{13}$ EMU and $\approx$ 400 Oe coercivity, affording moderate magnetic sensitivity. Infrared optical conductivity of the probe's magnetic coating has enabled our unprecedented application of simultaneous MFM and SNOM imaging throughout the present work. The probe was magnetized along its tip axis, aligned at about 17 degrees to the sample normal (c-axis) direction during measurements, enabling an overall attractive magnetic interaction between our probe and homogeneously ferromagnetic regions of the strained LCMO film (easy axis in-plane along the orthorhombic *b*-axis); this is leveraged in Figs. 3&4 to indicate changes in moment of the FMM phase. Meanwhile, out-of-plane magnetic "fringe fields" associated with ferromagnetic domain boundaries present the strongest source of magnetic contrast in our MFM images.

**Photo-excitation of strained $La_{2/3}Ca_{1/3}MnO_3$ within the cryo-SNOM**

As in previous work[14], a 1 kHz Ti:sapphire regenerative amplifier system (Spitfire Ace, Spectra-Physics) was used for 1.5 eV photo-excitation of the sample. After removing an infrared beam-splitter used for interferometric detection of the SNOM imaging signal, a flip-mirror is used to deliver collimated 1.5 eV optical pulses of sub-50 fs duration through the SNOM optics where they transmit through an IR-transmissive ZnSe window into the vacuum chamber of the cryo-SNOM. Pulses transmitted through an identical ZnSe window were characterized by frequency-resolved optical gating, revealing elongation of pulses not exceeding 150 fs, sufficient to coherently photo-excite the FMM phase of strained LCMO.[14] Optical pulses are focused to regions of the sample surface via the same off-axis parabolic mirror that serves as objective optic for SNOM measurements. Three-axis piezoelectric actuation of the parabolic mirror enabled delivery of diffraction-limited photo-excitation to regions of the sample with sub-micron selectivity (schematically indicated by Fig. 1d). The photo-excited FMM phase presents visible optical contrast against the parent insulating phase, which enabled confirmation of successful photo-excitation through in situ optical microscopy of the sample surface (example optical images are presented in SI).

Corresponding author:
† am4734@columbia.edu

Incremental delivery of single optical pulses to the sample is achieved by first disabling the intracavity Pockels cell of the Ti:sapphire laser, halting output of pulses at 1 kHz. A trigger signal is then sent to momentarily enable the Pockels cell, thus controllably emitting a single optical pulse from the laser.

**Finite Element Modeling of Strain**

Fig. 2b presents the predicted distribution of sample strain in proximity to cracks in the LCMO film, such as those observed in Figs. 1a-c. Using the finite element solver FEniCS[56], these results were obtained by solving for local rank-2 stress and strain tensors $\sigma$ and $\varepsilon$ in the boundary-value problem associated with elastic equilibrium $\nabla \cdot \sigma = 0$, where $\sigma = K\colon \varepsilon$ and $K$ is the rank-4 stiffness tensor of an isotropic solid. Here we consider a quasi-2-dimensional slice through the depth of the LCMO film in the $bc$-plane and normal to a crack propagating parallel to the orthorhombic $a$-axis. Epitaxial compressive strain of 0.7% along the LCMO orthorhombic a-axis and a tensile strain of +0.85% along the $b$-axis[15] (the latter rendered in false color through the cross-sectioned LCMO) throughout the strained film. Accordingly, tensile $b$-axis strain $\varepsilon_{bb} = +0.0084$ is imposed at the film-substrate interface in accord with the film epitaxy,[15] whereas the stress-free boundary condition $\vec{b} \cdot \sigma = \vec{0}$ is imposed along the $b$-axis throughout the film's depth at the location of the crack. Meanwhile the stress-free surface-normal boundary condition $\vec{c} \cdot \sigma = \vec{0}$ is imposed at the film surface. Although the full position-dependent strain tensor is obtained by this solution, results in Fig. 2b are visualized in terms of the salient b-axis component of dilatation strain $\varepsilon_{bb}$.

**Density functional theory calculations**

Density functional theory (DFT) calculations were performed using Vienna Ab-initio Simulation Package (VASP). The projector augmented wave (PAW) method is used to treat the core and valence electrons using the following electronic configurations: $5s^2 5p^6 5d^1 6s^2$ for La, $3s^2 3p^6 4s^2$ for Ca, $3p^6 3d^5 4s^2$ for Mn, and $2s^2 2p^4$ for O. The revised Perdew-Burke-Ernzerhof (PBE) functionals for solids (PBE-sol) are used in our calculation. We account for electron correlation in the Mn $d$ and La $f$ shell, using the DFT+$U$ method with effective

Corresponding author:
† am4734@columbia.edu

Hubbard $U$ values of $U_{eff}$(Mn_d)=3 eV and $U_{eff}$(La_f)=6 eV. A 1x3x1 supercell is used and the A-site cations are ordered in the $a$-$b$ plane to simulate the $Mn^{3+}/Mn^{4+}$ charge ordering in LCMO. The Brillouin zone is sampled using a 6×2×6 Γ-centered Monkhorst-Pack $k$-point mesh and integrations are performed using Gaussian smearing with a width of 10 meV. For the structural optimization, the lattice constants and atomic positions were relaxed until the forces on the atoms are less than 10 meV $Å^{-1}$.

**Ginzburg-Landau theory of strain-coupled magnetic transitions in strained LCMO**

The Ginzburg-Landau theory considered in the main text is developed on the basis of coupled order parameters $M$ and $Q$, denoting the local ferromagnetic moment and amplitude of a multi-modal lattice distortion (dominated by increase in the $Q_2$ Jahn-Teller distortion), respectively. The adimensionalized free energy density $F_{Q,M}$ of the electronic subsystem of LCMO can be expressed as (see SI):

$$F_{Q,M}/F_0 = \tau_Q Q^2 + Q^4 + e\, Q^2 M^2 + c_0 \tau_M M^2 + M^4 + \kappa_Q |\nabla Q|^2 + \kappa_M |\nabla M|^2. \qquad \text{(Eq. 2)}$$

Here $\tau_Q = (T - T_Q)/T_Q$ and $\tau_M = (T - T_M)/T_M$ reflect thermal proximity to the respective Neel (taken as $T_Q$) and Curie ($T_M$) temperatures of the system. Meanwhile, $c_0 (\approx 1)$ is a dimensionless measure of energetic asymmetry for $Q \neq 0$ and $M \neq 0$ phases, and $e$ denotes the strength of magneto-elastic coupling. Coincident low temperature metastability of AFI and FM phases calls for a minimum magnetoelastic coupling $e > 2$, producing local minima in the magnetoelastic free energy (Fig. 3a) corresponding to these phases. In the scenario where $T_M > T_Q$, Eq. 2 implies a thermal IMT into the FM phase, reflecting the behavior of bulk LCMO. Values for parameters are used corresponding to those in ref. [14]. "Domain wall" energies are quantified by charge order and ferromagnetic "stiffness" coefficients $\kappa_{Q,M}$.

Meanwhile, our ab initio calculations imply that coherent epitaxial strain along the film $b$-axis can further stabilize the AFI state by promoting the amplitude of $Q_2$ Jahn-Teller distortion. This motivates a contribution $F_\varepsilon$ to the overall free energy density arising from symmetry-allowed coupling between the Jahn-Teller distortion and strain:[31,57]

$$F_\varepsilon = \frac{1}{2}\sigma_f : \varepsilon_f = \frac{1}{2}\left(\varepsilon + \varepsilon_{\text{ep}} - \delta\varepsilon\, Q^2\right) : K : \left(\varepsilon + \varepsilon_{\text{ep}} - \delta\varepsilon\, Q^2\right). \quad \text{(Eq. 3)}$$

Here $\sigma_f$ and $\varepsilon_f$ are the local stress and strain tensors of the LCMO film, respectively, and $K$ is its stiffness tensor. Moreover, $\varepsilon$ denotes the strain tensor relative to the equilibrium elastic

Corresponding author:
† am4734@columbia.edu

configuration of the substrate (imparting the epitaxial constrain that $\langle\varepsilon_{\alpha\alpha}\rangle$=0 for in-plane directions $\alpha = a, b$ averaged over the substrate surface) and $\varepsilon_{\rm ep}$ denotes the epitaxial strain on the film, a diagonal tensor with non-zero in-plane values associated with those shown by arrows in Fig. 2b.[15] Lastly, $\delta\varepsilon$ is the "spontaneous strain"[31] tensor denoting the relative difference in relaxed (unclamped) LCMO lattice constants between AFI and FM phases at experimentally relevant temperatures ($T\sim$70K). This spontaneous strain was found to be significant (~1%) for over-doped LCMO upon entry to the insulating state, growing even larger for $T<T_M$.[12] Expanding Eq. 3 to lowest order in $Q^2$ establishes Eq. 1 of the main text. See SI for further details of the calculations implementing Eqs. 2&3 to obtain domain patterns as exemplified in Fig. 3b.

**Application of local stress by the AFM probe**

Results presented in Fig. 5 relied on local application of compressive stress from the same AFM probe utilized for SNOM and MFM imaging. Compressive stresses of 125, 250, and 500 MPa were applied by operating the microscope in contact-mode AFM feedback with feedback-controlled deflections of the probe cantilever (spring constant ≈ 2 N/m) of 12, 25, and 50 nm, resulting in net surface-normal-directed forces between of ≈ 25, 50, and 100 nN, respectively, between probe and sample. The values of compressive stress achieved at these forces are estimated by Hertzian contact mechanics,[58] taking account for elastic properties of the silicon probe and the sample together with the probe's radius of curvature. Compressive stresses quoted in the main text are depth averaged through the film thickness below the probe-sample contact area, whereas peak stresses are quoted from the probe-sample contact point. More details are provided in SI. After this protocol of compressive "erasure" is applied to the photoinduced FM phase, SNOM and MFM imaging of the same sample region are conducted after restoring the microscope to non-contact mode AFM feedback.

Corresponding author:
† am4734@columbia.edu

Corresponding author:
† am4734@columbia.edu

Corresponding author:
† am4734@columbia.edu

Corresponding author:
† am4734@columbia.edu

Corresponding author:
† am4734@columbia.edu

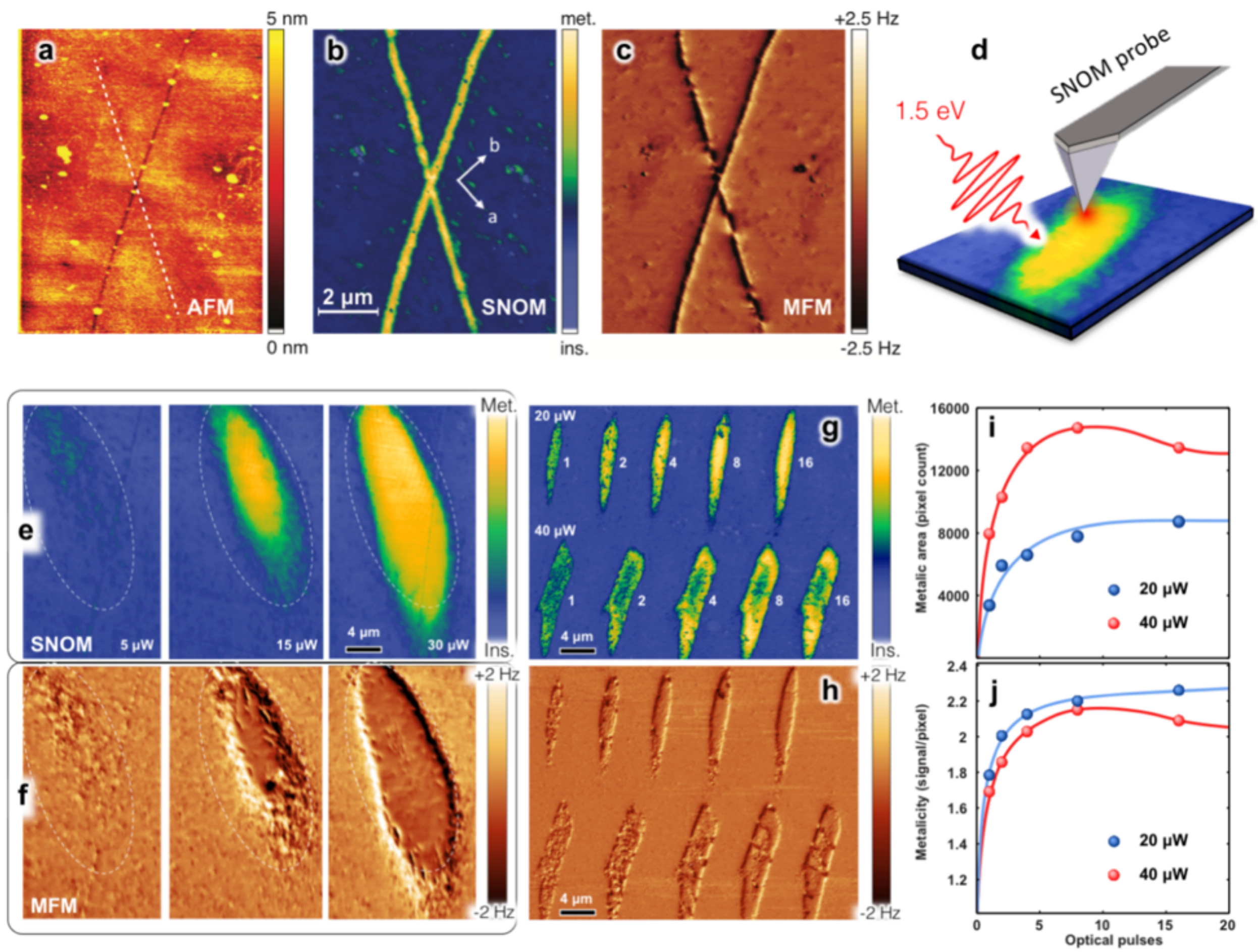

**Figure 1 | Nano-imaging of photoinduced ferromagnetic metal in epitaxial $La_{2/3}Ca_{1/3}MnO_3$ (LCMO).** a) Sample topography acquired by atomic force microscopy (AFM) revealing a crack in the LCMO film, as a well as a subtle step feature identified as a secondary crack (dashed white line). b) SNOM imaging at $T$=96K reveals metallicity localized within 100 nm of these cracks (crystallographic axes indicated), together with c) an in-plane magnetic moment observed by magnetic force microscopy (MFM), detected by shifts in resonance frequency of the cantilevered MFM probe. Otherwise faint ferromagnetic metal puddles are observed scattered through the bulk of the film. d) Schematic depiction of photo-excitation of LCMO within the cryogenic scanning near-field optical microscope. e) SNOM imaging of metallicity photoinduced at $T$=70K by a train of 1.5 eV optical pulses delivered from a Ti:Sapph laser at incremental fluence levels (5, 15, and 30 $\mu$W) to sample locations indicated by dashed white ovals. f) Maps of associated the local photoinduced magnetization acquired by simultaneous MFM, showing formation of single domain magnetism at fluences $\geq 15\,\mu$W. g) Local metallicity and h) magnetization of metallic domains photoinduced by incremental optical pulses delivered at two distinct fluence levels of 20 and 40 $\mu$W, shown in top and bottom rows, respectively; photoinduced domains are labeled by the number of cumulative pulses. i) Dependence of the overall metallic domain area and j) average metallicity upon the number of delivered pulses, demonstrating

Corresponding author:
† am4734@columbia.edu

cumulative growth up to a saturated size (after ∼16 pulses) that scales with fluence; 40 $\mu$W photo-excitation impairs metallicity due to thermal reentry of the insulating phase (see text).

Corresponding author:
† am4734@columbia.edu

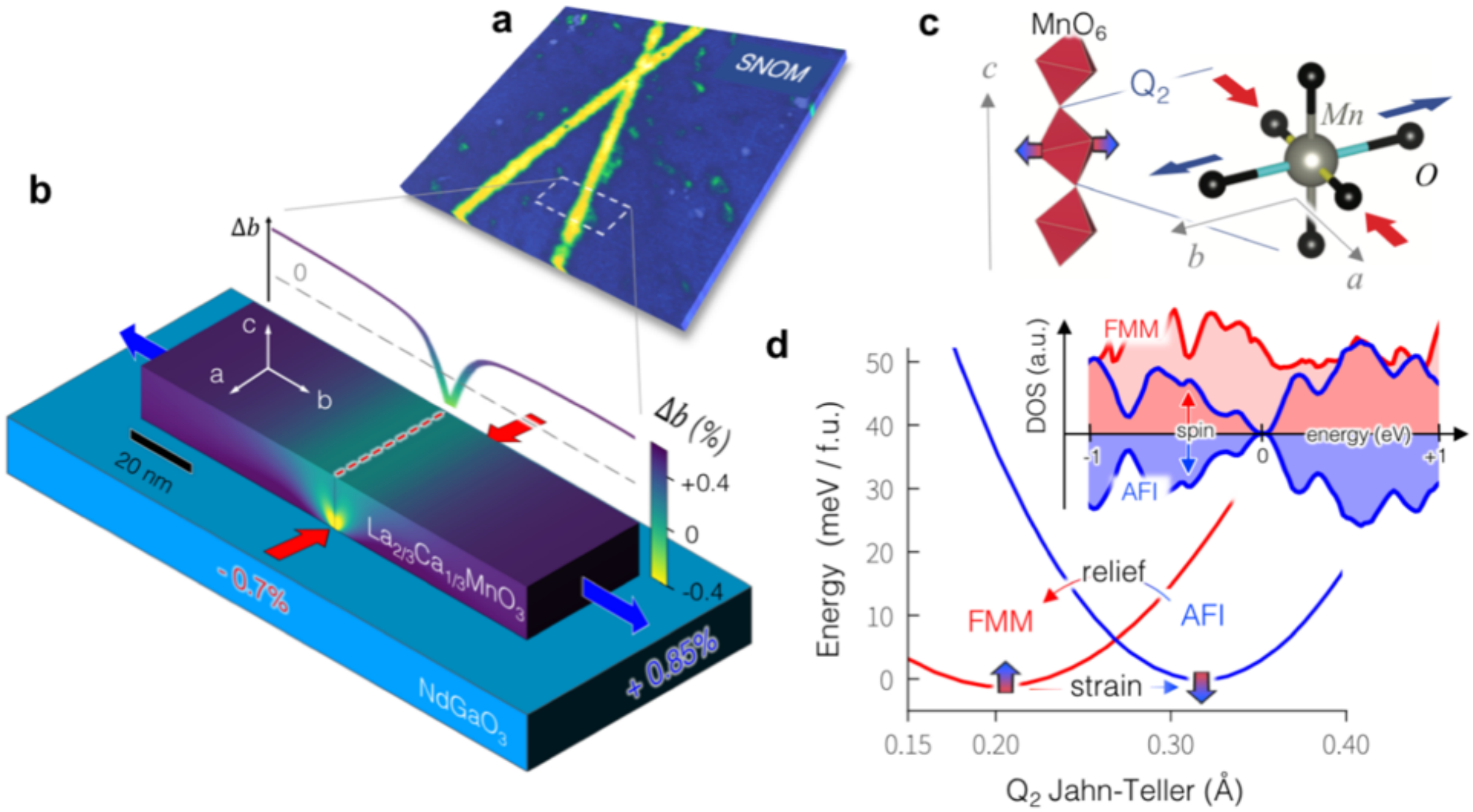

**Figure 2 | Strain-mediated suppression of hidden ferromagnetism.** a) The metallicity in proximity to cracks in the LCMO film (resolved by SNOM imaging) can be rationalized according to the local strain environment. b) Bi-axial strain normally causes reduction in the film's *c*-axis lattice constant, whereas strain relief at cracks (dashed white line) allows local restoration of an unstrained *c*-axis. Finite element simulation of the local c-axis strain (false color) predicts recovery of the lattice constant within a distance of the film thickness (26 nm) indicated by the depth-averaged strain profile (colored curve). c) Biaxial epitaxial strain is expected to induce $Q_2$ Jahn-Teller distortion of $MnO_6$ octahedra. d) Ab initio DFT+U calculations predict a critical epitaxial *b*-axis strain where a metastable antiferromagnetic insulator (AFI, blue) and ferromagnetic metal (FMM, red) phases attain competing energetic minima at distinct amplitudes of $Q_2$ distortion. Strain application or relief modulates the relative energies of these states as shown schematically. Inset: computed spin-resolved density of states (DOS) for FMM and AFI phases at the critical strain; note the >100 meV insulating gap in the AFI DOS.

Corresponding author:
† am4734@columbia.edu

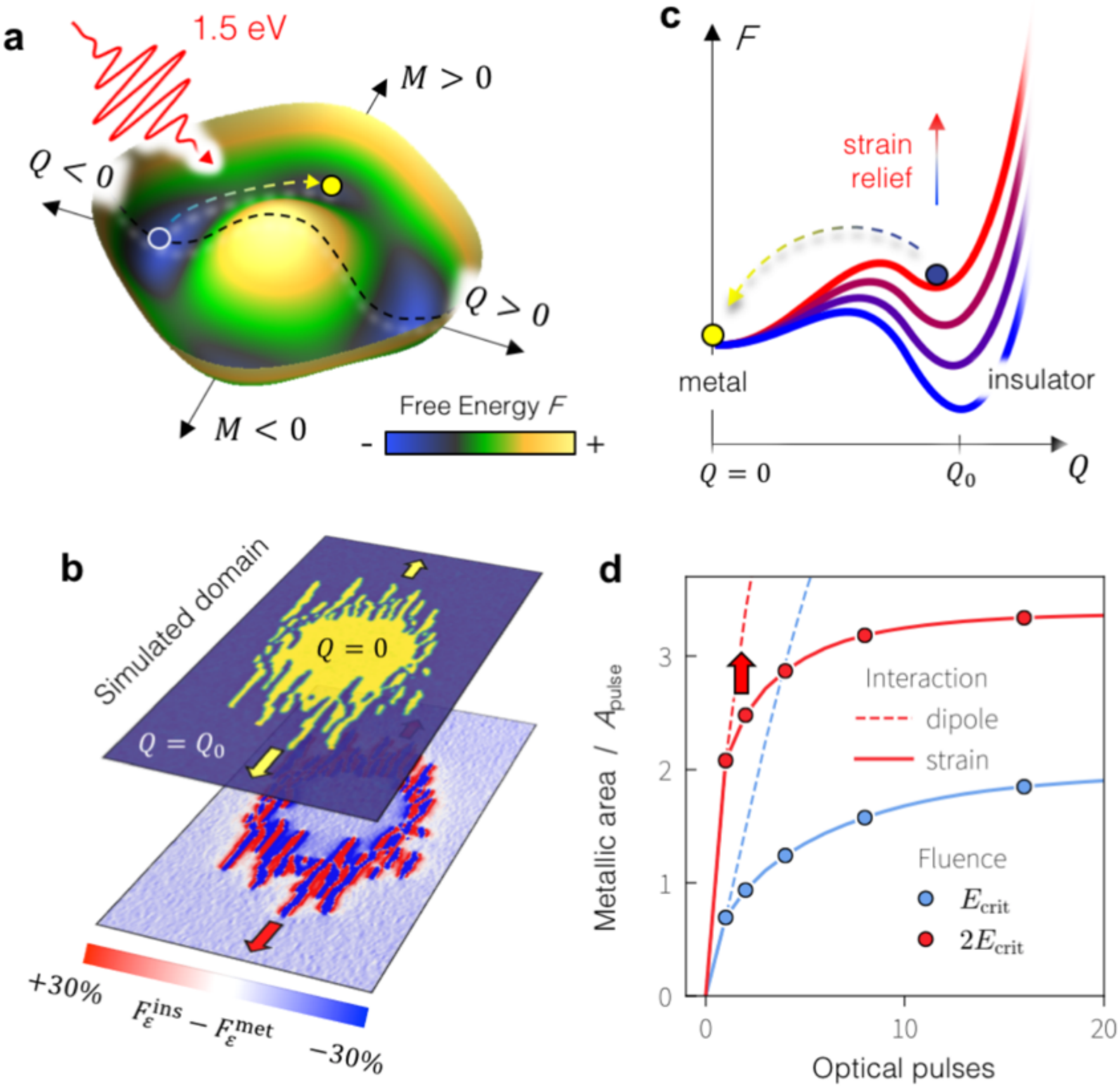

**Figure 3 | Co-active growth of photoinduced ferromagnetic metallic domains.** a) Schematic Landau free energy surface showing basins of stability corresponding to AFI (in blue, $Q \neq 0$) and metastable FMM (in yellow, $M \neq 0$) phases; 1.5 eV optical excitation of suitable fluence overcomes the energetic barrier (dashed arrow) to nucleate FMM. b) Phase-field modeling of an FMM domain (top); the resultant self-consistent strain field (bottom) lowers the elastic energy barrier to subsequent FMM photo-excitation at domain edges (red regions, see text). c) Relief of epitaxial strain co-actively reduces energy barriers for transitions into the FMM state, locally reducing the fluence threshold for photoinduced switching; curve colors correspond to spatial regions in the lower image of panel b). d) Simulations of co-active domain growth under incremental photo-excitation at fluence levels equal to and at twice the threshold field $E_{crit}$ for initial nucleation of an FMM domain. Solid and dashed lines consider alternatively relevant long-range energetics originating from accommodation strain and the FMM domain magnetic dipole field, respectively; the former shows superior agreement to observations (Fig. 1i)

Corresponding author:
† am4734@columbia.edu

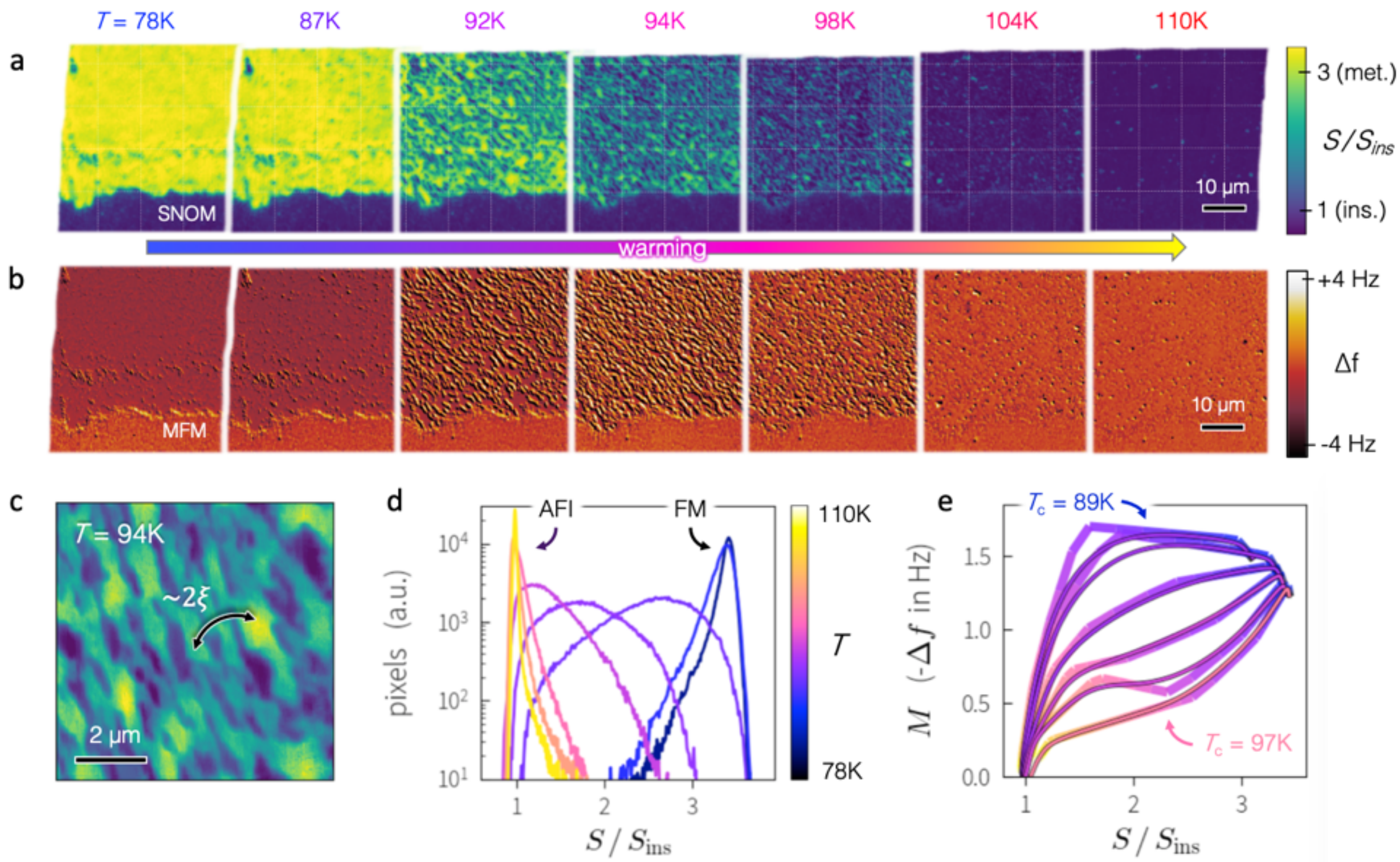

**Figure 4 | Thermal melting of the photoinduced ferromagnetic metal.** a) Thermal evolution of metallicity imaged by SNOM (signal $S$ in false color relative to that $S_{ins}$ of the insulator) in a large (~100x100 microns) FMM domain photoinduced at 70K; phase coexistence with the reentrant insulating phase sets in at temperatures $T$>90K, whereas the FMM phase vanishes almost completely by 110K. b) Simultaneously acquired MFM map of magnetism associated with the melting transition. c) SNOM image at $T$=94K detailing the striped pattern of phase coexistence; emergence of a characteristic periodic length scale $\xi$ is the hallmark of a strain-coupled insulator-metal transition (see text). d) Histogram representation of temperature-dependent SNOM signals $S$ from panel a), showing a weakly bimodal distribution distinguishing metal $(S > 1)$ from insulator $(S \approx 1)$ but with a continuous reduction in metallicity with $T$ (color-coded) approaching the melting transition. e) Simultaneous thermal evolution of $S$ and magnetization $M$ presented by several parametric curves associated with characteristic regions of the sample presenting "critical" temperatures $T_c$ (defined by a half-reduction in metallicity, see text) ranging from 89K to 97K; sample temperatures are color-coded as in d). The scaling of magnetic moment with metallicity is strongly inhomogeneous.

Corresponding author:
† am4734@columbia.edu

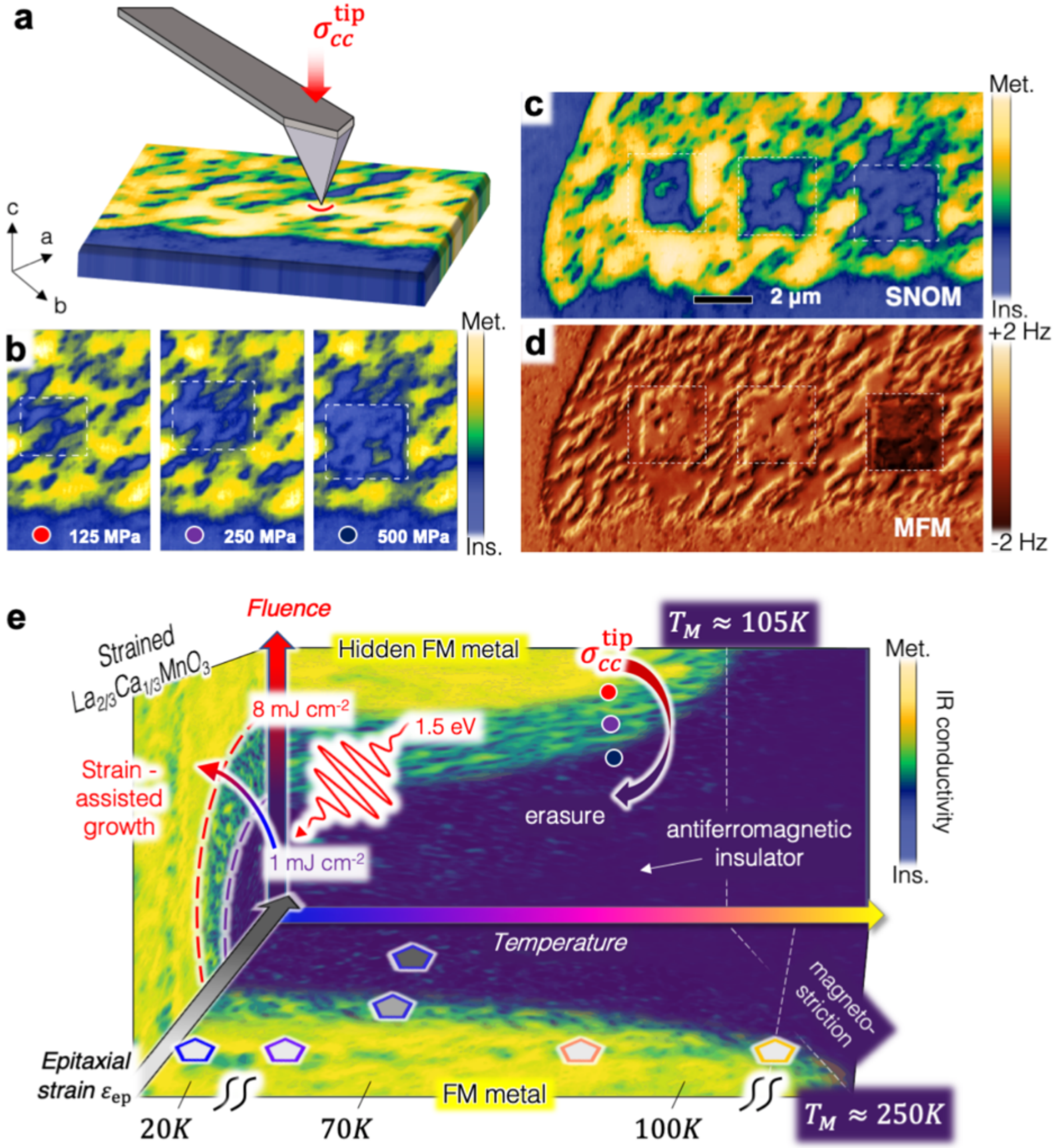

**Figure 5 | Nano-scale erasure of photoinduced metallicity**. a) Schematic depiction of compressive *c*-axis stress $\sigma_{cc}^{tip}$ delivered from the pointed tip of our cantilevered AFM probe to a local region in an extended domain of photo-excited FMM poised at *T*=90K. b) Metallicity of the FMM domain (detected by SNOM) after incremental application of local stress $\sigma_{cc}^{tip}$ = 125, 250, and 500 MPa, "written" by the AFM probe within a 3 micron-sized square (dashed white region). c) Stress-induced "erasure" of the FMM phase within three distinct regions (dashed white square) of the FMM domain by application of 500 MPa, demonstrating on-demand restoration of the insulating phase and d) associated demagnetization of the film

Corresponding author:
† am4734@columbia.edu

revealed by simultaneous MFM imaging. e) Phase diagram for strained $La_{2/3}Ca_{1/3}MnO_3$ showing stabilization of the insulating state (dark blue) at increasing in-plane epitaxial strain. Gray-shaded pentagons indicate strains/temperatures at which additional LCMO films were characterized by nano-imaging (see SI) without photo-excitation. Photoexcitation at 1.5 eV exceeding 8 mJ $cm^{-2}$ restores the hidden ferromagnetic metal (bright yellow), whereas co-active strain (blue/red arrow; see text) softens the switching threshold. Near the hidden metal's Curie temperature $T_M$, c-axis stress (e.g. $\sigma_{cc}^{tip}$) locally restores insulating end-states indicated by the colored circles coinciding with labels in b).

Corresponding author:
† am4734@columbia.edu

# Supplementary Information:
# Multi-messenger nano-probes of hidden magnetism in a strained manganite

A. S. McLeod [1,2,1] †, J. Zhang [1,*], M. Q. Gu [3], F. Jin [4], G. Zhang [1], K. W. Post [1], X. G. Zhao[5],

A. J. Millis [2], W. Wu [4,6], J. M. Rondinelli [3], R. D. Averitt [1], & D. N. Basov [1,2]

[1] Department of
Physics, University of California San Diego, La Jolla, California 92093, USA

[2] Department of Physics, Columbia University 538 West 120th Street, New
York, New York 10027, USA

[3] Department of Materials Science and Engineering, Northwestern University,
Evanston, Illinois 60208, USA

[4] Hefei National Laboratory for Physical Sciences at the Microscale, University of
Science and Technology of China, Hefei 230026, China

[5] Department of Mechanical Engineering, Boston University, Boston, Massachusetts
02215, USA

[6] Institute of Physical Science and Information Technology, Anhui University, Hefei
230026, China

Table of Contents

[1] These authors contributed equally to the present work.

# 1 Thermal behavior of strain-relieved ferromagnetic metal

Fig. 1 of the main text presents observation of ferromagnetic metal regions localized to incidental cracks in our coherently strained $La_{2/3}Ca_{1/3}MnO_3$ films. As described in the main text (Fig. 2), intentionally imposed epitaxial strain within the film is presumptively relieved in close proximity to these cracks. Images in Fig. S1 reveal the thermal behavior of metallicity and ferromagnetism as resolved by our temperature-dependent nano-IR and magnetic force imaging of the same region of the sample. Fig. S1a&b present line-scans of metallicity and magnetic signal, respectively, across a select crack region to illustrate their systematic temperature dependence. A metallic nano-IR response is found to emerge in proximity to these cracks at temperatures below 200K coinciding with onset of a monotonically increasing magnetic moment as temperature is reduced to the lowest probed temperatures (e.g. 50K). The 200K onset temperature for this strain-relieved ferromagnetic metal is relatively similar to the reported Curie temperature for the bulk ferromagnetic metal in unstrained $La_{2/3}Ca_{1/3}MnO_3$, substantiating the claim in the main text that its emergence in our samples can be ascribed to localized nano-scale restoration of the "natural" unstrained state of the $La_{2/3}Ca_{1/3}MnO_3$ system.

# 2 Magnetic phases in strained LCMO computed by DFT+U

The main text reports antiferromagnetic insulator (AFI) and ferromagnetic metal (FMM) phases for strained $La_{2/3}Ca_{1/3}MnO_3$ as computed by density functional theory (Vienna Ab-initio Simulation Package, or VASP), accounting for on-site Coulomb repulsion in the Mn *d* and La *f* shells (DFT+U). Other details of the calculation method are provided in Methods of the main text. Structural refinement (energy minimization) was performed for these two phases at varying levels of tensile *b*-axis strain up to +8%. As presented in Table 1, these calculations reveal that i) $MnO_3$ octahedral distortions develop monotonically with increasing strain primarily through a multi-modal distortion composed of $Q_2$ and $Q_3$ Jahn-Teller distortions, and ii) as highlighted in Fig. S2, the relative energy difference $\Delta E = E_{\mathrm{AFM}} - E_{\mathrm{FM}}$ between these two phases becomes negative at sufficiently large *b*-axis tensile strain (between +6 and +8% strain), suggesting that the AFI phase becomes the thermodynamically accessible ground state at appropriate levels of coherent epitaxial strain. These quantities are deduced from averaging the structural refinement throughout the stoichiometric unit cell used for calculation. Most critically, $Q_2$ and $Q_3$ Jahn-Teller distortions are uniformly larger in the AFI phase than in the FMM phase, and such relatively high levels of octahedral distortion preferably minimize elastic energy at sufficiently high tensile *b*-axis strain.

| | $Q_2$ distortion (Å) | | $Q_3$ distortion (Å) | | Energy (meV) | | |
|---|---|---|---|---|---|---|---|
| **Strain (%)** | **AFI** | **FMM** | **AFI** | **FMM** | **AFI** | **FMM** | **$\Delta E$ (meV)** |
| **+0** | 0.1358 | 0.0426 | -0.0482 | -0.0282 | -486.604 | -487.147 | 0.54358 |
| **+2** | 0.161 | 0.0402 | -0.1022 | -0.0833 | -486.554 | -486.920 | 0.36574 |
| **+4** | 0.1904 | 0.103 | -0.1547 | -0.1335 | -486.196 | -486.390 | 0.19447 |
| **+6** | 0.2223 | 0.19544 | -0.2051 | -0.1834 | -485.609 | -485.693 | 0.08343 |
| **+8** | 0.3169 | 0.3056 | -0.2505 | -0.2345 | -484.991 | -484.860 | -0.13098 |

## *2.1 Table 1 | Magnetic phases in strained LCMO computed by DFT+U*

Results of structural refinement of the $La_{2/3}Ca_{1/3}MnO_3$ lattice are deduced by averaging throughout the stoichiometric supercell used for the calculations. Levels of $Q_2$ and $Q_3$ Jahn-Teller distortion as well as energy per formula unit cell are compared for the antiferromagnetic insulator (AFI) and ferromagnetic metal (FM) phases at varying levels of fixed tensile *b*-axis strain up to +8%.

# 3 Landau description of the strain-coupled phase transition

Here we elaborate on the Landau theory proposed in the main text and moreover the pursuant simulations of the strain-coupled insulator-metal transition in strained $La_{2/3}Ca_{1/3}MnO_3$. First, as in Ref. [1], we advance a Ginzburg-Landau theory developed on the basis of coupled order parameters $M$ and $Q$, denoting the local ferromagnetic moment and amplitude of the $Q_2$ Jahn-Teller distortion, respectively. We regard the $Q^2 = 0, M^2 > 0$ as a ferromagnetic metal (FMM phase), and the $Q^2 = 0, M^2 > 0$ as an antiferromagnetic insulator (AFI phase). We describe the Landau free energy $f$ of the system by:

$$f = \int_{\Omega_f} dV \left(F_{Q,M} + F_\varepsilon^f\right) + \int_{\Omega_{\text{subs}}} dV\, F_\varepsilon^{\text{subs}} \qquad \text{(Eq. S1)}$$

with $F_{Q,M} = A\,\tau_Q Q^2 + B\,Q^4 + E\,Q^2M^2 + C\,\tau_M M^2 + D\,M^4 + \kappa_Q|\nabla Q|^2 + \kappa_M|\nabla M|^2$,

$F_\varepsilon^f \equiv \frac{1}{2}\sigma_f : \varepsilon_f = \frac{1}{2}\left(\varepsilon + \varepsilon_{\text{ep}} - \delta\varepsilon\, Q^2\right): K: \left(\varepsilon + \varepsilon_{\text{ep}} - \delta\varepsilon\, Q^2\right)$,

and $F_\varepsilon^{\text{subs}} \equiv \frac{1}{2}\sigma : \varepsilon$.

Here $\Omega_f$ and $\Omega_{\text{subs}}$ denote integration volumes extending throughout the LCMO film and the NGO substrate, respectively. Meanwhile, $\tau_Q = (T - T_Q)/T_Q$ and $\tau_M = (T - T_M)/T_M$ reflect thermal proximity to the respective Neel ($T_Q$) and Curie ($T_M$) temperatures of the system, whereas A, B, C, and D are energetic constants, and $E$ denotes the strength of magneto-elastic coupling. "Domain wall" energies are quantified by magneto-elastic "stiffness" coefficients $\kappa_{Q,M}$. The elastic energy density of the substrate $F_\varepsilon^{\text{subs}}$ is described through local second-rank stress and strain tensors $\varepsilon$ and $\sigma$, respectively, which relate to the local strain tensor $\varepsilon_f$ of the LCMO film through $\sigma_f = K\varepsilon_f$ and $\varepsilon_f = \varepsilon + \varepsilon_{\text{ep}} - \delta\varepsilon\, Q^2$. Here $K$ is a fourth rank stiffness tensor characteristic of the film and substrate, whereas $\varepsilon_{\text{ep}}$ denotes the epitaxial strain on the film, a diagonal tensor with non-zero in-plane values associated with the arrows shown schematically in Fig. 2b of the main text.[2] For simplicity in the present application, we consider LCMO and the NGO substrate as elastically isotropic solids for which elements of the stiffness tensor are given by $K_{ijkl} = \Lambda\delta_{ij}\delta_{kl} + 2\mu\frac{1}{2}(\delta_{ik}\delta_{jl} + \delta_{il}\delta_{jk})$, where $\delta_{ij}$ is the Kronecker delta symbol, and $\Lambda$ and $\mu$ are the material's Lamé and shear moduli, respectively.[3] Lastly, $\delta\varepsilon$ is the "spontaneous strain"[4] tensor denoting the relative difference in relaxed (unclamped) LCMO lattice constants between AFI and FMM phases at experimentally relevant temperatures ($T{\sim}70$K). This spontaneous strain was found to be significant (~1%) for over-doped LCMO upon entry to the insulating state, growing even larger for $T<T_M$.[5] Minimization of $F_\varepsilon^{\rm subs}$ implies an elastic energy in the film $F_\varepsilon^f$ characterized by $\varepsilon_f = \delta\varepsilon\, Q^2 - \varepsilon_{\rm ep}$, which is best minimized in the film volume by $Q^2 > 0$ when the tensors $\varepsilon_{\rm ep}$ and $\delta\varepsilon$ are congruent, as in the experimental scenario.

### *3.1 Adimensionalization of the Landau Free Energy*

For the purposes of utilization in simulation, we adimensionalize the free energy density $F_{Q,M}$ in Eq. S1 as follows. Using scale factors $q$ and $m$ we first define dimensionless order parameters $\bar{Q} \equiv Q \cdot q$ and $\bar{M} \equiv M \cdot m$, thereby setting a characteristic energy scale $F_0 \equiv b/q^4$ for which we have:

$$\frac{F_{Q,M}}{F_0} = \begin{array}{c} \tau_Q \frac{A\,q^2}{B} \bar{Q}^2 + \bar{Q}^4 + \frac{E\,q^2}{B\,m^2}\, \bar{Q}^2 \bar{M}^2 + \tau_M \frac{C\,q^4}{B\,m^2} \bar{M}^2 + \frac{D\,q^4}{B\,m^4} \bar{M}^4 \\ + \frac{q^4 \kappa_Q}{B} |\nabla \bar{Q}|^2 + \frac{q^4 \kappa_M}{B} |\nabla \bar{M}|^2 \end{array}.$$

We now freely relatively scale $q$ and $m$ according to $\frac{D\,q^4}{B\,m^4} \equiv 1$, whereby we can define the rescaled prefactors $e \equiv \frac{E\,q^2}{B\,m^2} = \frac{E}{\sqrt{B\,D}}$, $a \equiv \frac{A\,q^2}{B}$, and $c \equiv \frac{C\,q^4}{B\,m^2}$. Setting $a \equiv 1$ defines a scale $q \equiv \sqrt{B/_A}$ whereby $m \equiv \sqrt[1/4]{DB}/\sqrt{A}$ and $c = \sqrt{B/D} \cdot C/A$. Without loss of generality we subsume the dimensional prefactors on the gradient terms into the definitions of $\kappa_{Q,M}$. We thus obtain the adimensionalized free energy density presented in the manuscript:

$$F_{Q,M}/F_0 = \tau_Q \bar{Q}^2 + \bar{Q}^4 + e\, \bar{Q}^2 \bar{M}^2 + c_0 \tau_M \bar{M}^2 + \bar{M}^4 + \kappa_Q |\nabla \bar{Q}|^2 + \kappa_M |\nabla \bar{M}|^2. \qquad \text{(Eq. S2)}$$

In the manuscript and in what follows, we hereafter suppress the “bar” notation on adimensionalized order parameters. Moreover, although we do not show it here, $\bar{M}$ can be completely eliminated from the Eq. S2 by minimizing $F_{Q,M}$ with respect to $\bar{M}$ and obtaining $M = \bar{M}_{eq}(\bar{Q}, \tau_Q, \tau_M)$, whereby $F_{Q,M}$ can likewise be expressed strictly as a function of $\bar{Q}, \tau_Q$ and $\tau_M$, and for $e > 2$ takes the familiar form of a “double well” potential (see main text).

### *3.2 Minimization of Elastic Energy with respect to the elastic strain field*

We now consider minimization of the elastic energy terms $\int_{\Omega_f} dV\, F_\varepsilon^f + \int_{\Omega_{\rm subs}} dV\, F_\varepsilon^{\rm subs}$ with respect to the strain fields to remove them from explicit consideration. It will be shown that this minimization renders an effective long-range interaction between inhomogeneities in $Q^2(\vec{r})$ mediated by strain with a tendency to energetically favor coexistence of FMM and AFI phases with a striped nano-texture. For what follows, we consider a system comprising the thin LCMO film of thickness $t$ upon a substrate, here considered as an elastic half-space, with configuration and coordinate axes labeled as according to Fig. S3. First, we note that the substrate strain field is expressed in terms of the substrate lattice displacement field $\vec{u}$ by $\varepsilon(\vec{r}) \equiv \nabla\vec{u}(\vec{r})$, which implies the energy minimizing condition of elastic equilibrium $\nabla \cdot \sigma = \vec{0}$ within $\Omega_{\text{subs}}$. With zero net in-plane strain $\langle\varepsilon(\vec{r})\rangle_{\Omega_{\text{subs}}} = 0$ throughout the substrate, the only elastic energy contribution is given by $F_\varepsilon^{\text{subs}} = \frac{1}{2}\int_{z=0} dA\,(\hat{z}\cdot\sigma)\cdot\vec{u}$. Meanwhile, expanding the product in $F_\varepsilon^f$ develops a similar contribution:

$$\int_{\Omega_f} dV\,\frac{1}{2}\sigma_f : \varepsilon_f = \frac{1}{2}\begin{bmatrix} \int_{z=t} dA\,(\hat{z}\cdot\sigma_f)\cdot\vec{u} - \int_{\Omega_f} dV\,(\nabla\cdot\sigma_f)\cdot\vec{u} - \int_{z=0} dA\,(\hat{z}\cdot\sigma_f)\cdot\vec{u} \\ + \int_{\Omega_f} dV\,(\varepsilon_{\text{ep}} - \delta\varepsilon\, Q^2) : K : (\varepsilon_{\text{ep}} - \delta\varepsilon\, Q^2) \\ +2\int_{\Omega_f} dV\,\varepsilon : K : (\varepsilon_{\text{ep}} - \delta\varepsilon\, Q^2) \end{bmatrix}. \quad \text{(Eq. S3)}$$

Minimization of elastic energy terms on the first row in square brackets implies the surface boundary condition $\hat{z}\cdot\sigma_f = 0$ at $z = t$ and the volumetric equilibrium condition $\nabla\cdot\sigma_f = 0$ in $\Omega_f$. Combining $F_\varepsilon^f$ with $F_\varepsilon^{\text{subs}}$ cancels the boundary term at $z = 0$ and, meanwhile, integrand terms on the second row within square brackets proportional to $\delta\varepsilon^2 Q^4$ can be subsumed into $F_{Q,M}$ without any further regard. Subject to the equilibrium conditions, the last term in square brackets taken with other terms proportional to $Q^2$ thus comprise the total elastic energy density $F_\varepsilon$:

$$\int_{\Omega_f} dV\,F_\varepsilon = \begin{matrix} \left(\int_{z=t} dA - \int_{z=0} dA\right)\hat{z}\cdot(\sigma_{\text{ep}} - \delta\sigma\, Q^2)\cdot\vec{u} \\ -\int_{\Omega_f} dV\,\nabla\cdot(\delta\sigma\, Q^2)\cdot\vec{u} - \int_{\Omega_f} dV\,Q^2\,\delta\sigma : \varepsilon_{\text{ep}}. \end{matrix} \quad \text{(Eq. S4)}$$

Here $\sigma_{\text{ep}}$ and $\delta\sigma$ are shorthand for their strain counterparts under pre-multiplication by the fourth rank stiffness tensor $K$. We identify the present elastic energy integrand with that quoted in the manuscript: $F_\varepsilon \approx -Q^2\,\delta\varepsilon : K : (\varepsilon + \varepsilon_{\text{ep}})$. At the level of this Landau formalism, the last term of Eq. S4 proportional to $\delta\sigma : \varepsilon_{\text{ep}}$ is equivalent to an epitaxial strain-induced increase of the critical temperature for the AFI phase as discussed in the manuscript, and is unimpactful for the character of phase coexistence and domain formation; we thus exclude this term from present consideration.

The equilibrium condition $\nabla \cdot \sigma_f = 0$ within $\Omega_f$ implies that $\nabla \cdot \sigma = \nabla \cdot (\delta\sigma\, Q^2)$; thus, a local accommodation stress is naturally generated to minimize film elastic energy at the location of spatial inhomogeneities (e.g. domain walls) of the lattice distortion $Q$. This defines a linear boundary value problem for $\vec{u}$ whose solution can be obtained in terms of Green's dyadic functions $\underline{G}_g$ and $\underline{G}_T$ as:

$$\vec{u}(\vec{r}) \equiv \vec{u}_g + \vec{u}_T + \underline{\varepsilon}_{\parallel}\vec{r}_{\parallel} + \underline{\varepsilon}_{\perp}\vec{r}_{\perp},$$

$$\text{with } \vec{u}_g = \int_V dV'\, \underline{G}_g(\vec{r}-\vec{r}')\big(-\nabla\cdot\delta\sigma\, Q^2(\vec{r}')\big)$$

$$\text{and } \vec{u}_T = \int_{z'=0} dS'\, \underline{G}_T(\vec{r}-\vec{r}')(-\delta\sigma_z\, Q^2(\vec{r}')). \qquad \text{(Eq. S5)}$$

In the above, $\underline{\varepsilon}_{\parallel}$ and $\underline{\varepsilon}_{\perp}$ are constants of integration that can in principle accommodate, respectively, a net in-plane or out-of-plane expansion or contraction of the LCMO lattice (e.g. by an amount equal to the spontaneous strain $\delta\varepsilon$) for the removal of elastic energy under a transition from FMM to AFI; however finite $\underline{\varepsilon}_{\parallel}$ is prohibited by elastic energy cost in the substrate and its absence has been presupposed for the present derivation; consequently only oscillatory internal displacements $\vec{u}(\vec{r})$ are permitted. Meanwhile, without any experimental evidence for finite out-of-plane strain $\underline{\varepsilon}_{\perp}$ contributing to a reduction in elastic energy, we omit this term from consideration. Here the Green's functions $\underline{G}_g$ and $\underline{G}_T$ solve Navier's equation[3] for the elastic semi-infinite half-space in the presence of a point-like ("delta Dirac") body force $\vec{g}$ and surface traction $\vec{T}$, respectively, at a location $\vec{r}'$ and satisfy:

$$\nabla\cdot\big(K : \nabla\, \underline{G}_{g,T}\, \hat{g}\big) = \delta(\vec{r}-\vec{r}')\cdot\begin{cases}\hat{g} \text{ for } \underline{G}_g \\ \vec{0} \text{ for } \underline{G}_T\end{cases} \text{ for any } \hat{g}, \text{ and } \vec{r}, \vec{r}' \text{ in } \Omega_f,$$

$$\text{and } \hat{z}\cdot\big(K : \nabla\, \underline{G}_{g,T}\, \hat{T}\big) = \delta(\vec{r}-\vec{r}')\cdot\begin{cases}\vec{0} \text{ for } \underline{G}_g \\ \hat{T} \text{ for } \underline{G}_T\end{cases} \text{ for any } \hat{T}, \text{ and } \vec{r}, \vec{r}' \text{ at } z=t. \qquad \text{(Eq. S6)}$$

Insertion of the Eq. S5 solution into Eq. S4 reveals that volumetric elastic energy density $F_\varepsilon(\vec{r})$ effectively consists of long-range strain-mediated interactions between Eshelby-type inclusions[4] distributed according to the profile of $Q^2(\vec{r})$:

$$F_\varepsilon(\vec{r}) = -\frac{1}{2}\begin{bmatrix} \int_{\Omega_f} dV'\ \nabla\cdot\big(\delta\sigma\, Q^2(\vec{r})\big)\, \underline{G}_g(\vec{r}-\vec{r}')\, \nabla\cdot\big(\delta\sigma\, Q^2(\vec{r}')\big) \\ + 2\big(\int_{z'=0} dS' + \int_{z'=t} dS'\big)\, \nabla\cdot\big(\delta\sigma\, Q^2(\vec{r})\big)\, \underline{G}_T(\vec{r}-\vec{r}')\, \hat{z}\cdot\delta\sigma\, Q^2(\vec{r}') \end{bmatrix} \quad \text{(Eq. S7)}$$

To obtain the doubled second term within square brackets, we have applied the fact that $\underline{G}_T(\vec{r}-\vec{r}') = \underline{G}_T(\vec{r}'-\vec{r})$. For completeness, we also denote by $\partial F_\varepsilon(\vec{r})$ the elastic energy per unit surface area associated exclusively with the surface of the film at $z = t$ and the film-substrate interface at $z = 0$:

$$\partial F_\varepsilon(\vec{r}) = -\frac{1}{2}\int_{z'=t} dS'\,(\hat{z}\cdot\delta\sigma)\,Q^2(\vec{r})\,\underline{G}_T(\vec{r}-\vec{r}')\,(\hat{z}\cdot\delta\sigma)\,Q^2(\vec{r}') \quad \text{(Eq. S8)}$$

Because $\underline{G}_T(\vec{r}-\vec{r}')$ is long-range, this contribution has the qualitative effect of offsetting the aforementioned volumetric term proportional to $\delta\varepsilon^2 Q^4$; therefore we give $\partial F_\varepsilon(\vec{r})$ no further consideration for present consideration of the energetics driving pattern formation. Thus, Eq. S7 demonstrates that the effect of elastic strain in an insulator-metal transition associated with a spontaneous strain is to render an effective long-range interaction in the order parameter that leads to inhomogeneous accommodation strain and nontrivial domain morphologies, as demonstrated in the following. For the purposes of numerical computations discussed later we adimensionalize $F_\varepsilon$ by dividing through $F_0$; this simply amounts to re-scaling the stiffness $K$ in $\delta\sigma = K{:}\,\delta\varepsilon$ to the energy scale $F_0$.

## *3.3 Phenomenology of the strain-induced striped texture.*

The main text provides a heuristic explanation for the formation of a striped texture of the order parameter within the regime of phase coexistence for a strain-coupled first-order transition, whereas here we provide a more detailed explanation anchored in the reformulation of elastic energies in terms of long-range strain-mediated interactions between "inclusions" of $Q^2(\vec{r})$ (Eq. S7). The immediately relevant contributions $F_{\text{stripe}}$ to the Landau free energy density are the following:

$$F_{\text{stripe}}(\vec{r}) = \frac{1}{2}\left[\kappa|\nabla Q^2|^2 - \int_{\Omega_f} dV'\;\nabla\big(\delta\sigma\,Q^2(\vec{r})\big)\underline{G}_g(\vec{r}-\vec{r}')\,\nabla\big(\delta\sigma\,Q^2(\vec{r}')\big)\right]. \quad \text{(Eq. S9)}$$

Eq. S9 should be interpreted as a competition between 1) the reduction of elastic energy through formation of abundant domain boundaries ($|\nabla Q^2| > 0$), and 2) the energetic penalty of order parameter "stiffness" $\kappa$.

If we strictly consider $z$-independent and spatially periodic forms for $Q^2(\vec{r})$, e.g. at in-plane wave-vector $\vec{k}$, then the free energy has the following form:

$$F_{\text{stripe}} = \frac{1}{2}\int dA\left[t\,\kappa|\nabla Q^2|^2 - \int_V dA'\;\nabla\big(\delta\sigma\,Q^2(\vec{r})\big)\langle\underline{G}\rangle_g(\vec{r}-\vec{r}')\,\nabla\big(\delta\sigma\,Q^2(\vec{r}')\big)\right] \quad \text{(Eq. S10)}$$

Here $t$ is the thickness of the film along its $c$-axis, and $\langle\underline{G}\rangle_g \equiv \int_0^t dz \int_0^t dz'\,\underline{G}_g$ represents an "effective" lateral interaction kernel. It suffices for the present discussion to consider the long-range behavior of $\langle\underline{G}\rangle_g$:

$$\langle \underline{G} \rangle_g(\vec{r}) \sim t^2 \frac{(1+\nu)}{2\pi\, E} \left[ \frac{(1-2\nu)}{|\vec{r}|} \begin{pmatrix} 1 & 0 \\ 0 & 1 \end{pmatrix} + \frac{2\nu}{|\vec{r}|^3} \begin{pmatrix} x^2 & xy \\ xy & y^2 \end{pmatrix} \right] \text{ for } |\vec{r}| \gg t, \text{with } \vec{r} = \begin{pmatrix} x \\ y \end{pmatrix}.$$

(Eq. S11)

Here $\nu$ is the Poisson's ratio of the film/substrate's elastic medium, and $E$ is the associated Young's modulus. The Fourier transform of Eq. S10 within the $xy$-plane yields the energy density of a single Fourier component $Q^2{}_{\vec{k}}$, which for sufficiently small $k \equiv |\vec{k}| \ll t^{-1}$ reads:

$$f_{\text{stripe}}(\vec{k}) \sim \frac{t}{2} \left[ \kappa\, k^2 - t\, \left|\delta\sigma\, \vec{k}\right|^2 \frac{1+\nu}{2\pi\, E\, k} \right] \left|Q^2{}_{\vec{k}}\right|^2. \qquad \text{(Eq. S12)}$$

Here we have used the fact that $(\nabla Q^2)_{-\vec{k}} = (\nabla Q^2)^*_{\vec{k}} = -i\vec{k}(Q^2)^*_{\vec{k}}$ for real-valued $Q^2(\vec{r})$, where $^*$ denotes the complex conjugate. Eq. S12 encourages reinterpretation of the energetic competition in Eq. S10 through an effective "scale dependent surface tension" for the order parameter equal to $\kappa_{\text{eff}}(k) \equiv \kappa - t\, |\delta\sigma|^2 \frac{1+\nu}{2\pi\, E\, k}$. In this case Eq. S12 is minimized for a very particular value of $k$ at equilibrium:

$$k_{eq} \sim \frac{1}{2\pi} \frac{1+\nu}{E\, \kappa} \left|\delta\vec{\sigma}_{x,y}\right|^2 \sim \frac{1}{2\pi(1+\nu)} \left(\frac{1-\nu}{1-2\nu}\right)^2 \frac{E\, t}{\kappa} \left(\frac{\Delta\alpha_{x,z}}{\alpha_{x,z}}\right)^2. \quad \text{(Eq. S13)}$$

Here $\delta\vec{\sigma}_{x,y}$ denotes the principle (maximal magnitude) component of $\delta\sigma$ projected onto the $xy$-plane, and $\Delta\alpha_{x,y}/\alpha_{x,y}$ denotes the change in unit cell size associated with the phase transition, projected along $\delta\vec{\sigma}_{x,y}$. This value of $k_{eq}$ defines an emergent length scale $\xi \equiv 2\pi/k_{eq}$ for the formation of striped phase coexistence. The orientation of $\vec{k}_{eq}$ is likewise expected to match that of $\delta\vec{\sigma}_{x,y}$, provided that this principle component is sufficiently large compared with the other principle components. When this criterion is not satisfied, further minimization of elastic energy gain be gained through emergence of more interesting equilibrium textures,[6] although we don't consider these in the present work. Most importantly, as observed with nano-imaging in the present case of strained $La_{2/3}Ca_{1/3}MnO_3$, this "principle" orientation is definitively the crystallographic $b$-axis.

This result accords with the length scale $\xi$ described in the main text and follows closely derivations for the scale of phase inhomogeneities derived for other strain-coupled phase transitions in equilibrium[6] . Physically speaking, Eq. S13 reflects the outcome of two competing real-space contributions to the system free energy. Long-range strain interactions (in proportion to $E$) tend to "thin" the length scale of inhomogeneities down to the scale of their depth $t$ so as to mitigate the elastic distortions they impart to the substrate, whereas short-range or microscopic domain-wall energetics tend to "coarsen" these inhomogeneities so as to reduce inter-phase interfacial energies. At the intrinsic equilibrium length scale $\xi$, these effects precisely cancel. In principle, nano-scale sensitivity to $\xi$ affords sensitivity to the energetics of both phenomena. Fig. 4c of the main text reveals the length scale as ≈ 800nm for the present sample of strained $La_{2/3}Ca_{1/3}MnO_3$.

While the phenomenology remains as stated above, the imprecise equality in Eq. S13 can be modified by a factor of order unity when considering that higher harmonics of $\vec{k}_{eq}$ arise in $Q^2(\vec{r})$ to mutually minimize local terms in the Landau free energy (such as the temperature-dependent "double-well" potential $F_{Q,M}(Q, M, T)$, not shown in Eq. S10). To accurately calculate the equilibrium configuration of $Q^2(\vec{r})$, such as that presented in Fig. S4 in comparison with experimental nano-IR data, we therefore use a numerical approach as described in the following section.

### *3.4 Minimization of Landau free energy*

Fig. 3b of the main text presents a simulated striped domain pattern of the $Q^2 = 0$ phase that results from minimization of the Eq. S1 Landau free energy starting from an initial circular domain configuration. Accompanying this simulated domain pattern is a strain field $\varepsilon(\vec{r})$ associated with the distribution $Q^2(\vec{r})$. This strain field is utilized to compute $F_\varepsilon^{\mathrm{met}}$ and $F_\varepsilon^{\mathrm{ins}}$, which are defined as the sum of terms $F_\varepsilon^f + F_\varepsilon^{\mathrm{subs}}$ from Eq. S1 evaluated respectively in the homogeneous $Q^2 = 0$ and $Q^2 = Q_0^2$ (the equilibrium value associated with the CO-AFI phase) states. Here we describe how the energy minimization obtaining striped domain patterns is implemented in practice. In thermal equilibrium, the spatial configuration adopted by the order parameter function $Q^2(\vec{r})$ is that which minimizes the Landau free energy $F$ presented in Eq. 1 of the main text. This configuration can be computed by solving the Euler-Lagrange differential equations, given the free energy density $f$ presented in Eq. S1:

$$\frac{\partial f}{\partial Q^2} - \nabla \cdot \frac{\partial f}{\partial \nabla Q^2} \equiv \mu(\vec{r}) = 0 \quad \text{for all } \vec{r} \text{ in } V. \qquad \text{(Eq. S13)}$$

Here $\mu(\vec{r})$ represents a local "chemical potential" for the system, expressions for which were derived accounting for the long-range strain energy contribution $f_\varepsilon(\vec{r})$ of Eq. S7; these will be detailed in forthcoming publication. We implement an iterative solution to Eq. S13 based on a Cahn-Allen model for kinetics of a non-conserved order parameter[7] :

$$\frac{dQ^2(\vec{r})}{dt} = -R\,\mu(\vec{r}). \qquad \text{(Eq. S14)}$$

Here $R$ is a (*ad hoc*) positive rate constant, and time-stepping from an initial "trial" configuration $Q^2_{\text{init}}(\vec{r})$ (which for Fig. 3b of the main text takes the form of an initially circular domain of the $Q^2 = 0$ FMM phase) accomplished by solving Eq. S14 via an implicit Crank-Nicholson scheme[8] at incremental times $t$, using the finite element solver FEniCS[9] evaluated at ever-increasing time-steps $\Delta t$ chosen to bring the system rapidly and arbitrarily close to the asymptotic condition $\mu(\vec{r}) = 0$. At this point the kinetics are concluded, resulting in an equilibrium configuration $Q^2(\vec{r})$ that (though perhaps locally / metastably) minimizes the global Landau free energy $F$ at a particular temperature $T$. Fig. S4b presents an example equilibrium configuration obtained as the result of this energy minimization at a temperature $T \approx T_Q^*$, where $T_Q^*$ denotes the transition temperature for $Q^2 > 0$ ordering renormalized by the epitaxial strain. The stripe periodicity $\xi$ is here selected (according to Eq. S12) to resemble the experimental data in Fig. S4a reproduced from the main text. Figs. S4c&d present the computed "static structure factors" associated with phase coexistence in the experimental and simulated images, respectively. Here the static structure factor of an image (that spatially resolves, as in our case, the order parameter associated with the phase transition) is obtained from the spatial Fourier transform of its auto-correlation.[10] Peaks in the static structure denote values of spatial momentum associated with modulated phase coexistence, here unambiguously identifying $\vec{k}_{eq}$ and thereby also $\xi$. The resemblance between experimental and simulated phase coexistence patterns and static structure factors attests to the underlying validity of the Landau theory presented in Eq. S1 as an explanation for the patterns of phase coexistence for the photoinduced phase transition in strained $La_{2/3}Ca_{1/3}MnO_3$.

# 4 Model for cumulative strain-assisted growth of photo-excited domains

Fig. 3d of the main text presents a simple simulation of the strain-assisted cumulative growth of ferromagnetic metal (FMM) domains induced by subsequent optical pulses delivered to the LCMO sample, as resolved experimentally in Figs. 1i&j. Building on the phenomenological Landau theory of Supplementary Section 3, here we present a simple model that rationalizes this finding. Without recourse to any particular microscopic model of the photoinduced insulator-metal transition, we first assume that a photoexcitation field exceeding a critical amplitude $E_{\mathrm{crit}}$ will be capable of activating the phase transition from antiferromagnetic insulator (AFI) to FMM. This corresponds to a critical number of $n_{\mathrm{crit}} = \alpha\, E_{\mathrm{crit}}^2/\hbar\omega$ of absorbed photons per unit area of the film, with $\alpha$ an optical absorption cross section for the AFI phase at $\hbar\omega = 1.5$ eV. In the context of our phenomenological Landau theory described in Supplementary Section 3, we can represent this "trigger" mechanism with an additional photoinduced term $F_{\mathrm{phot}} \sim \frac{n}{n_{\mathrm{crit}}} Q^2$ in the (adimensionalized) Landau free energy density. Here $n$ is the number of photons absorbed in the film per unit volume, and $Q$ is the order parameter associated with the multi-modal lattice distortion of the AFI phase. Thus, three terms of the Landau free energy density are proportional to $Q^2$ and together (their sum we call $F_{Q^2}$) contribute to the "balance" of the double-well free energy density, as shown schematically in Fig. 3c of the main text:

$$F_{Q^2} = F_\varepsilon + F_{\mathrm{phot}} \approx Q^2 \left[ \left( \delta\varepsilon : K : \varepsilon_{\mathrm{ep}} \right) + \left( -\, \delta\varepsilon : K : \varepsilon \right) + \frac{n}{n_{\mathrm{crit}}} \right]. \quad \text{(Eq. S15)}$$

For simplicity, we consider that all terms here are adimensionalized through division by the characteristic energy density $F_0$ (see Supplementary Sec. 3.2). The first term in brackets is negative in the homogeneous AFI phase, through which epitaxial strain in our LCMO film favors the AFI phase ground state as already discussed. The second term in brackets describes coupling to inhomogeneous strain, which is produced by insulator-metal domains as described and calculated according to Eq. S5, and shown graphically for a realistic FMM domain structure in Fig. 3b of the main text. In general, this term is positive immediately outside FMM domains due to the accommodation strain produced at FMM-AFI domain boundaries. The last term in brackets describes $F_{\mathrm{phot}}(\vec{r}) \propto E(\vec{r})^2$, proportional to the local intensity of the photoexcitation field. Eq. S15 makes clear that the effect of inhomogeneous strain $\varepsilon(\vec{r})$ is to effectively modulate the critical photoexcitation density $n_{\mathrm{crit}}$.

We denote the term within brackets of Eq. S15 as $\Delta F$, indicating the local difference in free energy density between insulator and metallic phases, and remark that a critical value $\Delta F_{\mathrm{crit}}$ of order unity in this model is sufficient to remove the kinetic energy barrier between AFI and FMM phases, thus destabilizing the $Q^2 > 0$ (AFI) phase and triggering the insulator-metal transition (cf. Fig. 3a&c of the main text). In Fig. S6 we plot each of the contributions to $\Delta F$ as well as their sum for photoexcitation under individual sequential optical pulses. Fig. S6a reveals that photoexcitation with a first optical pulse will nucleate an FMM domain whose size is determined foremost by the area within which $E(\vec{r}) > E_{\mathrm{crit}}$. In this simple model, this nucleated domain size can in principle be smaller than the full-width at half-maximum (focal size) of the optical excitation pulse. Fig. S6b reveals that the effect of a second optical pulse is determined in part by the self-consistent strain field of the already nucleated FMM domain; here for simplicity the blue curve is computed according to Eqs. S5&S15 for a quasi-circular FMM domain using values for the spontaneous strain $\delta\varepsilon$ corresponding to bulk LCMO taken from literature.[5] As shown by the magenta curve, the sum of these contributions to $\Delta F$ defines a new radius within which $\Delta F > \Delta F_{\mathrm{crit}}$, thus triggering growth of the pre-existing FMM domain. Fig. S6c&d indicate the same scenario after 3 and 4 optical pulses, respectively, showing co-active strain-assisted growth of the FMM domain. Fig. S6e generalizes Fig. 3c of the main text to resolve the phenomenological free energy landscape at individual spatial locations within the photoexcitation envelope. Fig. S6f summarizes the evolution of $\Delta F(r)$ upon each stage (individual optical pulse) of sequential photo-excitation, rationalizing co-active and strain-assisted growth of the FMM domain.

The process of cumulative FMM domain growth eventually saturates when $\Delta F_{\mathrm{crit}}$ cannot be exceeded at domain radii $r$ much exceeding that where $E(r) = E_{\mathrm{crit}}$. The spatial distribution of the realistically computed strain field (blue curves in Fig. S6b-d) combined with the gaussian photoexcitation envelope defines a cumulative growth process that generally saturates in this model after 4 or 5 optical pulses, in agreement with our experimental results (Fig. 1i&j of the main text). On the other hand, this behavior is inconsistent with the result obtained in our model when the blue curve is replaced by a power law curve such as $\Delta F \propto r^{-2}$, as for a free energy density modulated by a magnetic dipole field originating at the center of the FMM domain, to which we might expect coupling of the magnetic moment order parameter *M* (Eq. S1). In such a scenario, due to the long-range nature of such a putative power law, we find saturation of the FMM domain size only after an unrealistically large number of optical pulses, obtaining very large saturation sizes for the simulated FMM domain; these outcomes are reflected by the dashed curves in Fig. 3d of the main text. Together with this simple model, our findings thus substantiate the influence of localized strain fields on the dynamics of photoexcitation. More generally, these results demonstrate how coupling of relevant order parameters to strain can globally or locally modulate the energetic landscape of the insulator-metal transition in strained LCMO.

# 5 Mechanisms for nano-resolved erasure of the photoinduced ferromagnetic metal

The main text describes "erasure" of the photoinduced ferromagnetic metal (FMM) phase through compressive strain applied from the atomic force microscope probe operated in contact-mode AFM feedback, with feedback-controlled deflections of the probe cantilever. The result of this erasure for several probe-sample forces was shown in Fig. 5a-d of the main text. Coupling of the insulator-metal transition to strain in epitaxial $La_{2/3}Ca_{1/3}MnO_3$ implies that external stress could modify the free energy landscape dictating relative stability of FMM and antiferromagnetic insulator (AFI) phases, especially for temperatures not far from the Curie temperature $T_M \approx 105\text{K}$ we observe for photoinduced FMM domains. As reasoned in the main text, a normal stress $\sigma_{cc}^{tip}$ applied from the tip along the c-axis of the LCMO film is expected to produce in-plane tensile stress associated with the Poisson effect- namely, the tendency for any material to expand in directions perpendicular to the direction of compression. The resultant in-plane strain $\varepsilon_{\parallel}$ is expected to couple to the order parameter operative in the insulator-metal transition, as described in detail by the Landau theory of Supplementary Section 3. In particular, $\varepsilon_{\parallel}$ is expected to lower the phenomenological free energy barrier for transitions into the epitaxially favored AFI phase, provided that $\varepsilon_{\parallel}$ is sufficiently large (on the scale of the epitaxial strain $\varepsilon_{\text{ep}}\sim 1-2\%$). With a Poisson ratio of approximately 0.4 and Young's modulus of about 150 GPa for bulk LCMO, local stresses of order $\sigma_{cc}^{tip}\sim 1$ GPa might be expected to significantly impact the energetics of the system. This value is in accord with the minimum necessary pressure (500 MPa and above) inferred for complete erasure of the FMM phase, as described by Hertzian contact mechanics.

Hertzian contact mechanics can relate the contact radius $a$ to the force $F$ delivered onto a flat elastic half-space (with Poisson ratio $\upsilon_1$ and Young's modulus $E_1$) from an elastically deformed hemispherical probe of undistorted radius $R$ (with Poisson ratio $\upsilon_2$ and Young's modulus $E_2$):[11]

$$a = \sqrt[3]{\frac{3FR}{4}\left[\frac{1-\nu_1^2}{E_1} + \frac{1-\nu_2^2}{E_2}\right]}. \qquad \text{(Eq. S16)}$$

Here we can associate the elastic properties indexed by 1, 2 with those of our silicon probe and bulk LCMO, respectively, according to values reported in Table S2. We consider a probe radius of $R = 30$ nm suitable for description of our silicon probe with 20 nm thick cobalt-chromium coating (240AC-MA, Mikromasch USA). The local pressure is then approximated as $P_{\max} = F/(\pi a^2)$, decaying with distance $r$ into the half-space as $P(r) \approx P_{\max} a^2/(a^2 + r^2)$.[3] Accordingly, probe-sample forces as high as the 100 nN as applied in our experiments should achieve contact radii as large as 4 nm, peak pressures of 5.5 GPa at the contact point, and average pressures of 550 MPa within a sample volume equal to the film thickness ($t = 26$ nm) cubed.

Meanwhile, as mentioned in the main text, since the AFM probe in our microscope is held at elevated temperature compared with the sample, some local heating of the sample due to the probe contact may be expected during the protocol of "erasing" the photoinduced FMM phase. We analyze this possibility here in detail by applying a thermal model of conductive heat transfer from the probe to the sample as shown in Fig. S7, and using values of thermal conductivity $\kappa_{\mathrm{Si}}$ for the AFM probe body (silicon), $\kappa_{\mathrm{mag}}$ for the magnetic metallic coating on the AFM probe (magnetic Co-Cr alloy), $\kappa_{\mathrm{film}}$ for the film (approximated as "bulk" LCMO), and $\kappa_{\mathrm{subs}}$ for the film substrate ($NdGaO_3$, or "NGO"). With probe in contact with the sample, the top-left inset equations of Fig. S7 relate the temperature $T_{\mathrm{top}}$ at the surface of the LCMO film to the temperatures $T_{\mathrm{probe}} \approx 300\mathrm{K}$ of the probe and $T_{\mathrm{sample}} \approx 90\mathrm{K}$ of the sample by way of thermal resistances in the problem. We seek to address whether a simple thermal model can predict an average temperature increase within the sampling volume of the LCMO film above the nominal 90K sample temperature (measured at the substrate) towards the requisite temperature for thermal melting of the FMM phase, namely $T_M \approx 105\mathrm{K}$.

The thermal resistance of the probe is considered as the sum of thermal resistances from a silicon "slab" comprising the probe cantilever and silicon cone comprising the probe tip shaft, terminating at the contact radius $a$ following from Hertzian mechanics. Standard formulas for thermal resistance of these structures are given in Fig. S7, and relevant geometric parameters are given by Table S3. Moreover, the metallic coating of the AFM probe is expected to result in nano-scale roughness across the probe-sample contact area. The effective thermal resistance of the associated with nano-scale roughness at the contact junction has been investigated intensively through thermal transport measurements[12], scaling quantitatively as $R_{\text{junction}} = 1/mF$. Here $F$ is the force at the probe-sample junction and $m$ is an empirical constant measured as approximately 2 W (K N)$^{-1}$ for junction pressures much less than 1 GPa, falling to as low as 0.7 W (K N)$^{-1}$ for junction pressures exceeding 1 GPa, reflecting the "saturation" of atomistic contact point proliferation at the junction upon increasing pressure. In order to form a conservative estimate for thermal transport between our probe and the sample, we assume the higher value for $m$, and moreover we disregard the thermally resistive (compared to silicon) magnetic metallic coating on the AFM probe which should intervene at the thermal junction. Therefore, our following estimations for thermal transport are expected to overestimate the temperature rise on the sample due to thermal contact with the AFM probe.

Meanwhile, as expressed in the bottom-left inset of Fig. S7, the thermal resistance of the film and sample (collectively $R_{\text{sample}}$) might be estimated from the simple boundary value problem of heat transport given by $\nabla^2 T = 0$ solved in spherical coordinates on a semi-infinite thermal half-space of conductance $\kappa_{\text{sample}}$, with boundary conditions $T = T_{\text{top}}$ at radial coordinate $r = a$ and $T = T_{\text{sample}}$ at $r = \infty$ (namely, at the comparatively distant point of sample temperature measurement). Heat current through the sample is then evaluated through $\frac{dQ(r)}{dt} = -2\pi r^2 \kappa_{\text{sample}} \frac{dT}{dr}$, from which the thermal resistance is approximated as $R_{\text{sample}} \approx 1/(2\pi\kappa_{\text{sample}} a)$. However, this result fails to account for non-uniform thermal conductance of the sample comprising the LCMO film of thickness $t = 26$nm atop the NGO substrate, both with drastically dissimilar thermal conduction $\kappa_{\text{film}} \approx \kappa_{\text{subs}}$ according to Table S2. We therefore applied a finite element method[9] to solve the problem of inhomogeneous heat transport given by $\nabla \cdot (\kappa \, \nabla T) = 0$ in a quasi-infinite half space subject to the appropriate boundary conditions. Fig. S8a presents the simulation result in the form of a cross-sectional temperature distribution at the lateral coordinate $y = 0$, with the film occupying the space $z \in [0, t]$, and Fig. S8b displays the resultant temperature drop is predominantly across the sample thickness, credited to the comparatively low thermal conductance of LCMO. Computing again the resultant heat flow by $Q = -\kappa_{\text{film}} \int_{z=0} dx \, dy \ \frac{dT}{dz}$, we find an effective thermal resistance that is 2 to 3 times greater than the simple approximation for $R_{\text{sample}}$ evaluated with $\kappa_{\text{sample}} \equiv \kappa_{\text{subs}}$. Moreover, Fig. S9 displays the average film temperature increase expected within a cylindrical volume of radius $R$ centered at the probe-sample contact point, relative to the temperature difference $\Delta T = T_{\text{top}} - T_{\text{sample}}$. The characteristic size of a domain of "erased" FMM phase is given by $R \geq 50$nm, within which according to Fig. S9b we expect a temperature increase of about 3% of the total temperature difference $\Delta T$.

We combine the operative thermal resistances to estimate the temperature increase within a relevant volume of the LCMO film for several probe-sample forces used in the "erasure" experiment (graphics of Fig. 5a&b of the main text are reproduced as Fig. S10a-&b) as shown Fig. S10c. The solid and dashed curves associate with a variable probe-sample contact radius of 2-4 nm, respectively. We find that temperature increase in the sample is as high as a few degrees K for the largest probe-sample forces, with a maximum temperature increase of 6K in the limit of an infinitely thermally resistive sample material ($\kappa_{\text{sample}} \to 0$). We reiterate that these results disregard i) the several-fold decrease in $m$ associated with the thermal contact resistance expected for operative peak probe-sample pressures exceeding 1 GPa, and ii) the thermally resistive metallic coating on the magnetic probe, both of which are expected to appreciably reduce the predicted temperature increase within the sample volume. Although we find a plausible several degrees K increase in sample temperature associated with our "erasure" protocol, we do not establish sufficient evidence for a purely thermal mechanism for erasure of the FMM phase, which should require local heating to a temperature of at least $T_M \approx 105$K, which we find otherwise necessary for thermal melting of this photoinduced phase. We therefore conclude that a "thermally assisted" pressure-induced mechanism is the most plausible explanation for erasure of the FMM phase. Future investigations with a cryogenically cooled probe tip should further clarify the relative roles of temperature and strain for this mode of controllable nano-scale phase manipulation.

| | Young's modulus | Specific heat | Thermal conductivity | Density |
|---|---|---|---|---|
| Silicon | 130-185 GPa | 0.7 J/(g*K) @ 80 K | 1.5 W/(cm*K) (300 K) 20W/(cm*K) (80 K) | 2.32 g/cm$^3$ |
| $Co_{85}Cr_{15}$ | 210 GPa | 0.390 J/(g*K) | 0.094 W/(cm*K) | 10 g/cm$^3$ |
| LCMO | 143-175 GPa | 50 J/(mol*K) | 0.01 W/(cm*K) | 6.1 g/cm$^3$ |
| NGO | 238.5 GPa | 30 J/(mol*K) | 0.3 W/(cm*K) | 7.57 g/cm$^3$ |

### *5.1 Table S2 | Material properties related to probe-sample thermal conduction*

| $a$ | $W$ | $t$ | $L_1$ | $L_2$ | $D$ | $m$ | $F$ |
|---|---|---|---|---|---|---|---|
| ~2-4 nm | 28 microns | 3 microns | 225 microns | 15 microns | 15 microns | 0.7-2 W / (K N)* | 50-100 nN |

### *5.2 Table S3 | Mechanical parameters related to probe-sample thermal conduction*

* The empirical constant *m* associates with thermal resistance of the rough probe-sample contact junction.[12]

# 6 Nano-imaging of additional $La_{2/3}Ca_{1/3}MnO_3$ films at varied temperatures and epitaxial coherence

Figure S11 presents nano-imaging data acquired from other $La_{2/3}Ca_{1/3}MnO_3$ films grown at varying levels of epitaxially coherence to the substrate by way of their controlled annealing time (ranging from 3-10 hrs among the studied films) under an oxygenated environment.[2,13] These data inform the region of the phase diagram before photo-excitation as presented in Fig. 5 of the main text. Fig. 11a displays SNOM and MFM images of the same film considered in the main text, annealed for 10 hrs achieving a the dominantly antiferromagnetic insulating (AFI) low-temperature phase, acquired in a 5x5 micron area at $T$=80K. Sparse metallic "islands" oriented roughly perpendicular to the film $b$-axis are found distributed in the insulating background of the film, each with a detectable magnetic moment as shown by contrast in the MFM image (resolving local magnetic fields by the induced frequency shift $\Delta f$ of the magnetic probe cantilever).

By comparison, a film annealed for only 6 hrs in the same environment to achieve partial epitaxial strain coherency is found to exhibit insulator-metal phase coexistence in roughly equal proportion in an appreciable temperature range below 100K, as shown by characteristic SNOM and MFM images in Fig. S11b acquired in a 10x10 micron area at *T*=71K. The abundance of inhomogeneous ferromagnetic metallic (FMM) domains in this film is demonstrated by the detected metallic infrared (IR) conductivity together with ubiquitous but variable ferromagnetic moment across the sample surface. We attribute this magnetic texture to a fine-grained coexistence of AFI and FMM phases in roughly equal proportion, associated with an environment of quenched disorder within the film attributable to inhomogeneous (incoherent) epitaxially strain.

Finally, Fig. S11c presents a sequence of SNOM images acquired at temperatures from 23K to 195K probing the local optical conductivity of a $La_{2/3}Ca_{1/3}MnO_3$ film annealed for only 3 hrs under oxygenated atmosphere. Images here are presented in a 20x10 microns field of view including fiducial markers consisting of small gold islands (bright yellow in the images) sputtered onto the sample surface, enabling quantitative normalization of nano-infrared signals $S$ into the range $S \in [0,1]$ relative to the value acquired on the reference markers. These images clearly reveal the near-continuous emergence of dominantly metallic conductivity across the film surface for temperatures below approximately 180K. Moreover, resistive transport measurements of this film (not shown) reveal onset of metallic conductance below 150K, in accord with our nano-resolved infrared images. Although MFM images were not acquired on this film, we nevertheless attribute regions of high metallic conductivity with the FMM phase characteristic of "bulk" LCMO. Micron-scale "islands" of weaker (insulating) optical response are found to persist at low temperature (e.g. 60K) within the metallic background of this film, and we attribute these islands to a plausibly antiferromagnetic insulating phase whose stability is already "nucleated" by the incoherent epitaxial strain of the substrate. However, this discussion presents only a preliminary and phenomenological explanation for phase coexistence identified within this nearly "bulk-like" LCMO film, and further investigations are warranted for films grown with such incoherent (and perhaps strongly inhomogeneous) epitaxial strain. Regardless of the outstanding unresolved microscopic details, the nano-imaging results presented here for LCMO films prepared at varying levels of epitaxial strain coherency substantiate the part of the phase diagram presented in Fig. 5c of the main text pertinent to the thermodynamic “starting point” of these films before photoexcitation.

# 7 Supplementary References

# 8 Supplementary Figures

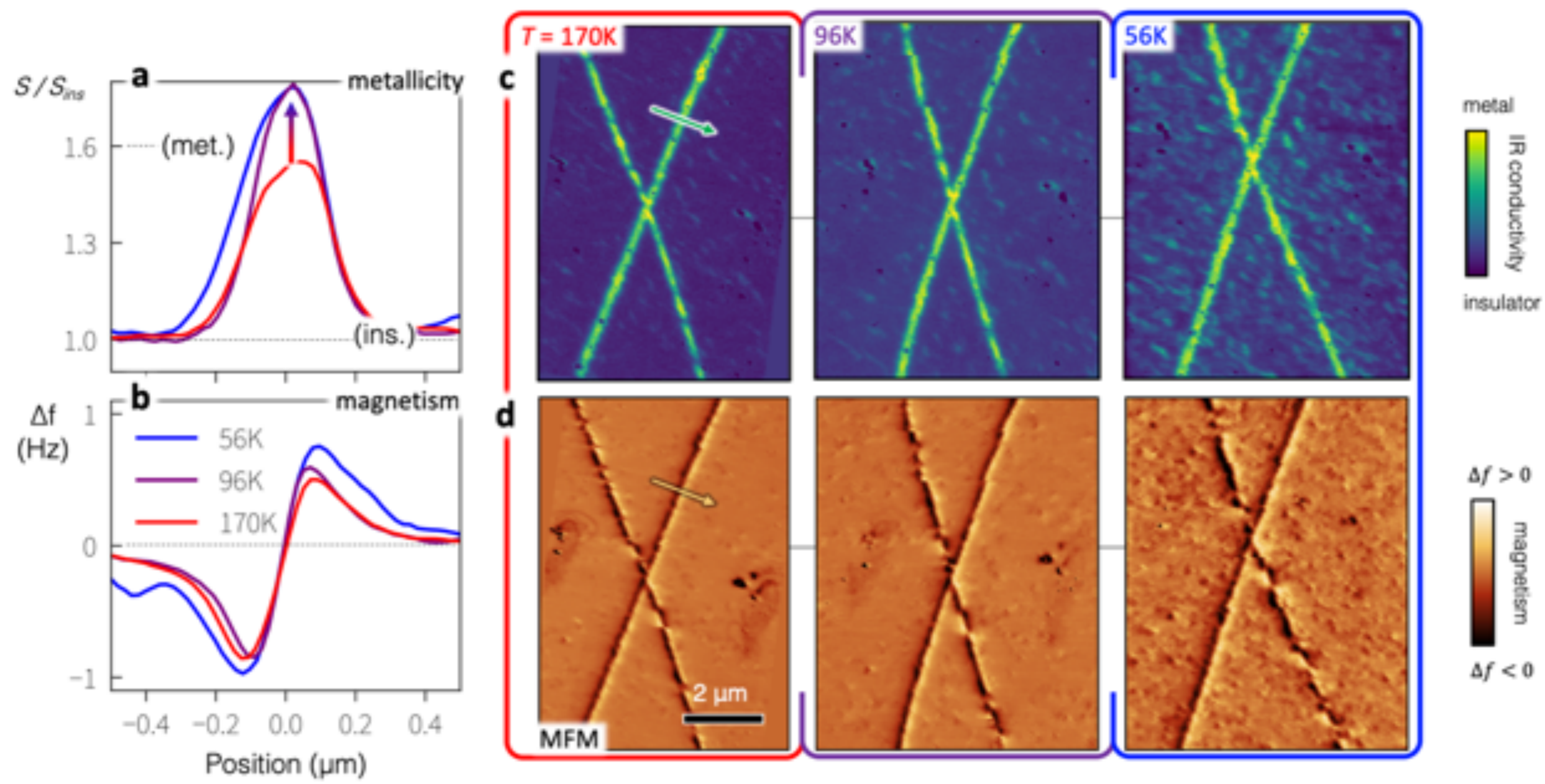

## 8.1 Figure S1 | Thermal evolution of the strain-relieved ferromagnetic metal.

**a)** Temperature dependence of the metallic response across a strain-relieved (cracked) region of the LCMO film as resolved through the nano-infrared signal $S$ relative to that of the insulating background $S_{ins}$. Colored curves associate with the legend in panel b). The metallicity is position resolved along the arrowed line traversing one of the cracks as shown in panel c). Metallicity onsets below $T$=200K and saturates at the center (position "zero") of the strain-relieved cracks at temperatures below 100K. **b)** Magnetic force microscopy signals from the strain-relieved metallic phase resolved by frequency shift of the magnetic cantilever across the same arrowed path in panel d). In-plane magnetization along the LCMO b-axis produces the antisymmetric magnetic line shape detected by our out-of-plane polarized magnetic probe. Magnetization likewise onsets below 200K and increases further to 56K. **c)** Low-temperature nano-infrared images associated with the line-traces in panel a), and **d)** magnetization images associated with those in panel b).

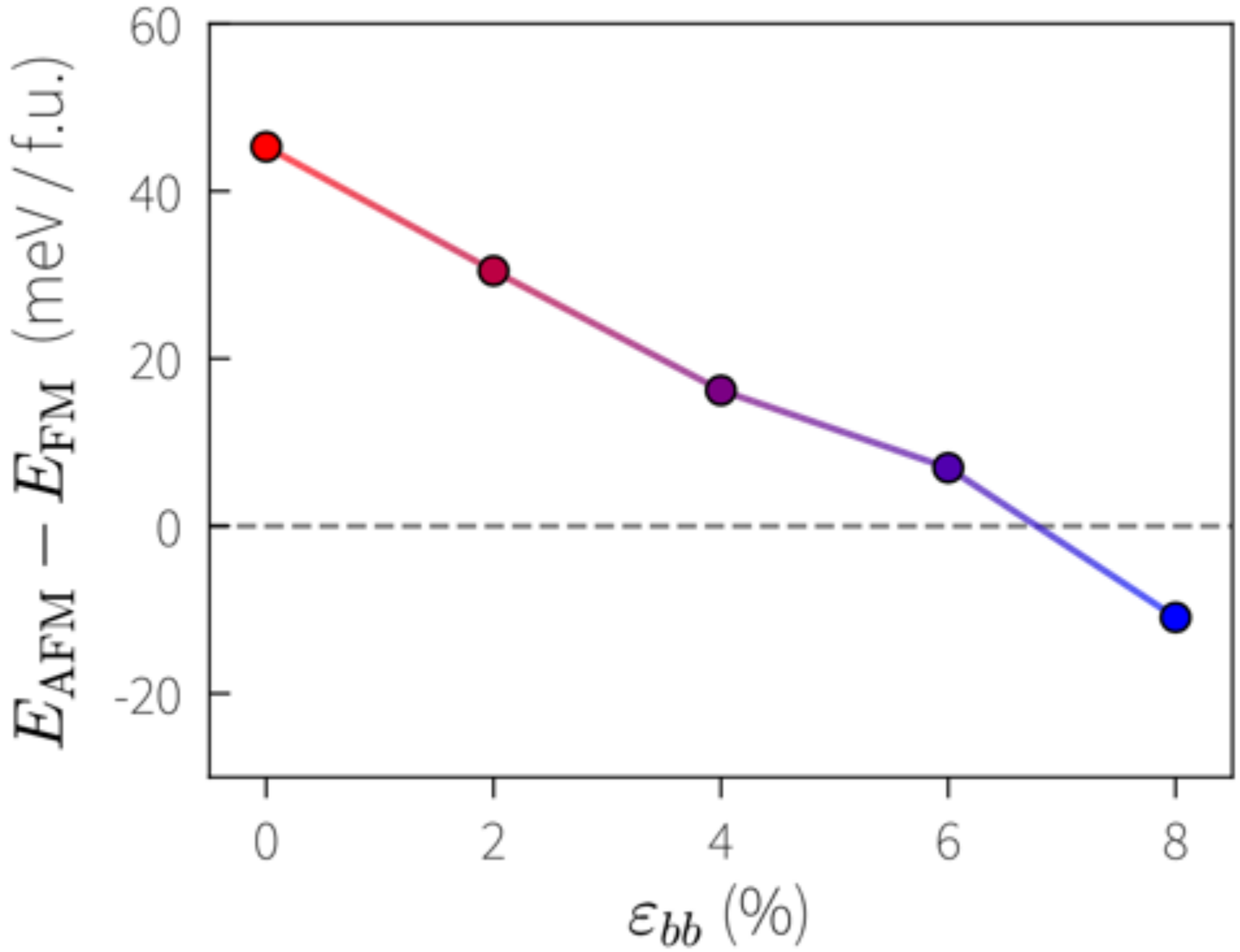

### 8.2 *Figure S2 | Relative stability of predicted ferromagnetic metal and antiferromagnetic insulator.*

Energy difference per formula unit between the antiferromagnetic (AFM) insulator and ferromagnetic (FM) metal phases computed by DFT+U at several values of tensile b-axis strain $\varepsilon_{bb}$ up to 8% applied to $La_{2/3}Ca_{1/3}MnO_3$. The antiferromagnetic insulator is stabilized as the ground state relative to the ferromagnetic metal at a critical strain in excess of the experimental value (~1% tensile strain along the b-axis) but within the same order of magnitude.

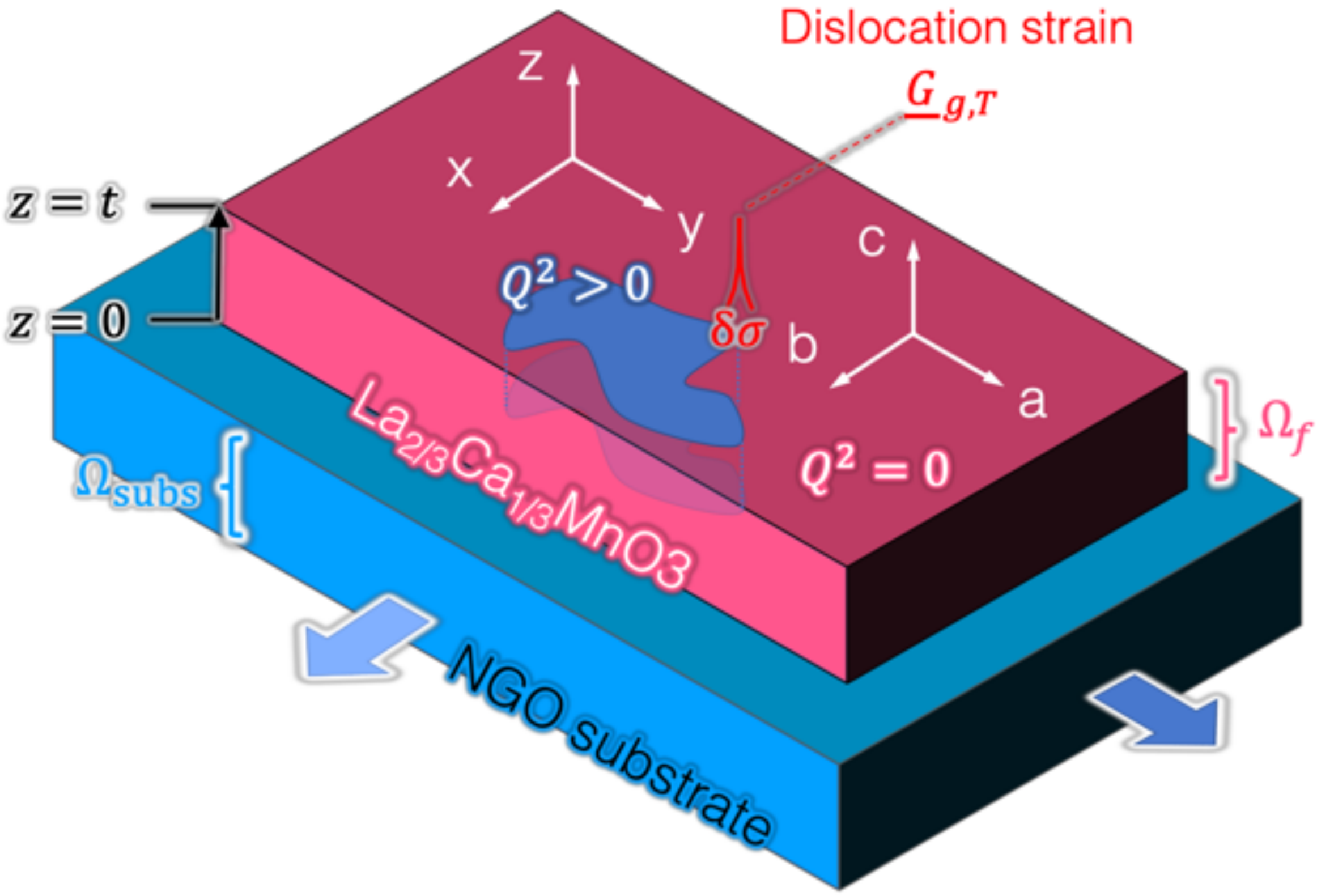

## 8.3 *Figure S3 | Schematic for strain-coupled Landau theory of epitaxial LCMO.*

Geometric configuration of the $La_{2/3}Ca_{1/3}MnO_3$ film epitaxially clamped to the $NdGaO_3$ (NGO) substrate. Coordinate axes and crystal axes of the film are as indicated. The film/substrate interface is located at $z = 0$ whereas the film surface is at $z = t$. Regions of the film in the $Q^2 > 0$ (charge-ordered antiferromagnetic insulator) configuration introduce a dislocation stress of magnitude $\delta\sigma$ at boundaries with the $Q^2 = 0$ configuration (ferromagnetic metal), originating strain fields throughout the film and substrate as described by the Green's dyadic functions $\underline{G}_g(\vec{r})$ and $\underline{G}_T(\vec{r})$.

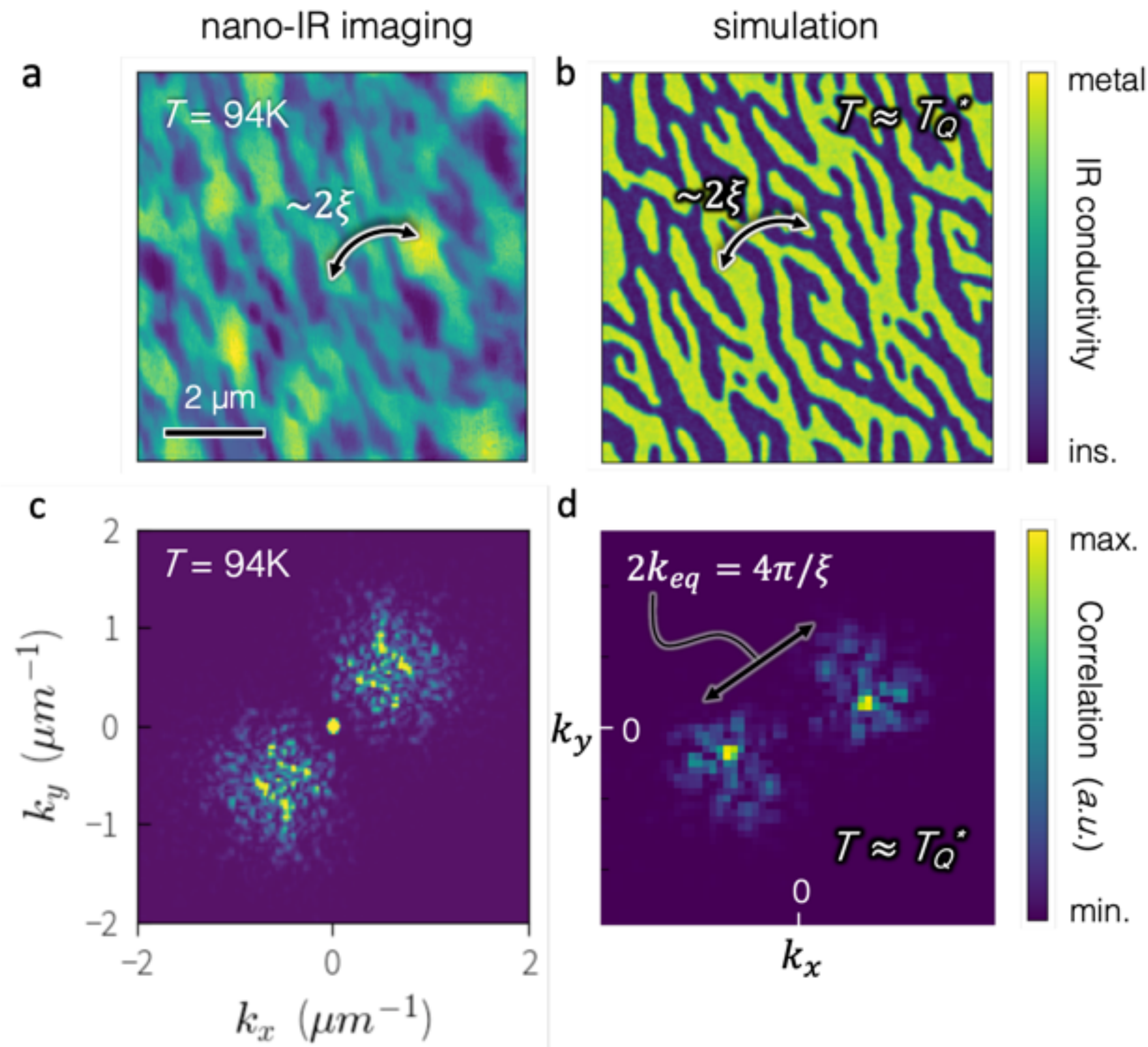

## 8.4 *Figure S4 | Spatial texture of insulator-metal phase coexistence in epitaxial LCMO.*

**a)** Nano-IR image of phase coexistence between metal (green/yellow) and insulator (dark blue) acquired at T=94K during the melting transition of the photoinduced ferromagnetic metal. Twice the periodicity $\xi$ for striped phase coexistence is indicated. **b)** Corresponding simulated pattern of phase coexistence resulting from energy minimization in the Landau theory of Eq. 1 for a temperature close to the $Q^2 > 0$ ordering temperature $T_Q^*$. **c)** & **d)** Static structure factors (see supplementary text) associated with the experimental and simulated phase coexistence images, respectively. Each shows the presence of two peaks in Fourier space denoting the equilibrium wave-vectors associated with striped phase coexistence.

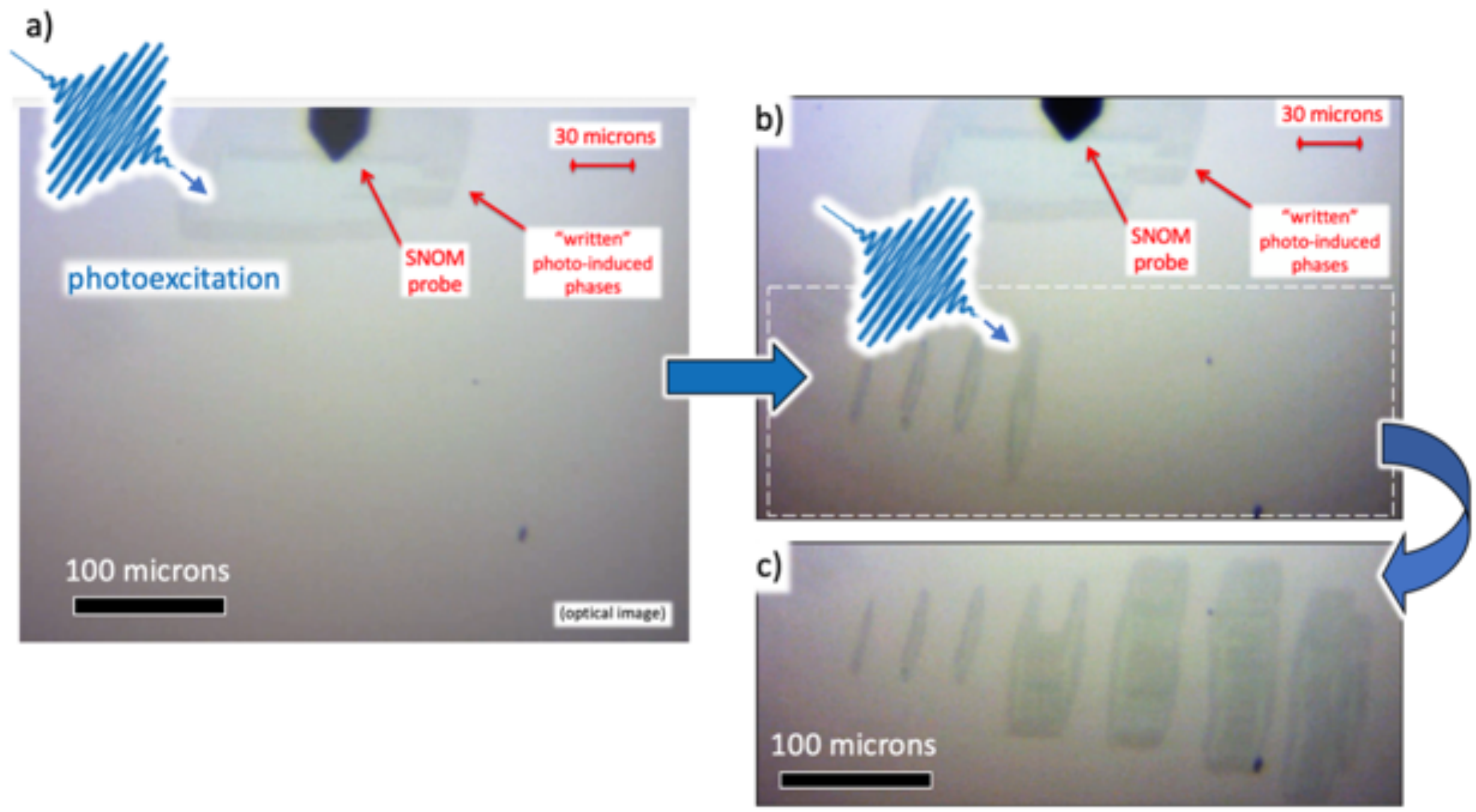

## 8.5 *Figure S5 | Optical images of the photoinduced ferromagnetic metal.*

Top-down visible microscopy views of the LCMO film surface inside the cryogenic scanning near-field optical microscope (cryoSNOM) after several rounds of selective photo-excitation of the hidden ferromagnetic metal, all at a sample temperature T=70K. The SNOM probe is visible at the top of the roughly 500x500 micron field of view. The focusing optic normally utilized for SNOM detection is instead here utilized to deliver 1.5 eV optical pulses to select regions of the sample surface by way of piezo-electric micro-positioning. **a)** Photo-excitation from continuously delivered 1.5 eV optical pulses at a fluence of 80 mJ cm$^{-2}$ serves to "write" a large patch of ferromagnetic metal under the SNOM probe, visible as a ~200 micron wide patch of darkened visible contrast, with a central region of disordered ferromagnetic metal (see main text) appearing with a slightly lighter optical contrast at its center. **b)** Examples of "single pulse" photo-excitation events at increasing fluences (left to right) of 30, 50, and 80 mJ cm$^{-2}$. At the right-most photo-excitation event (marked by the optical pulse), the excitation mode has been switched back to continuous application of optical pulses. **c)** This exposure mode is used to "write" the letters "U C S D" (for University of California San Diego) with visible contrast provided by the photo-excited ferromagnetic metal.

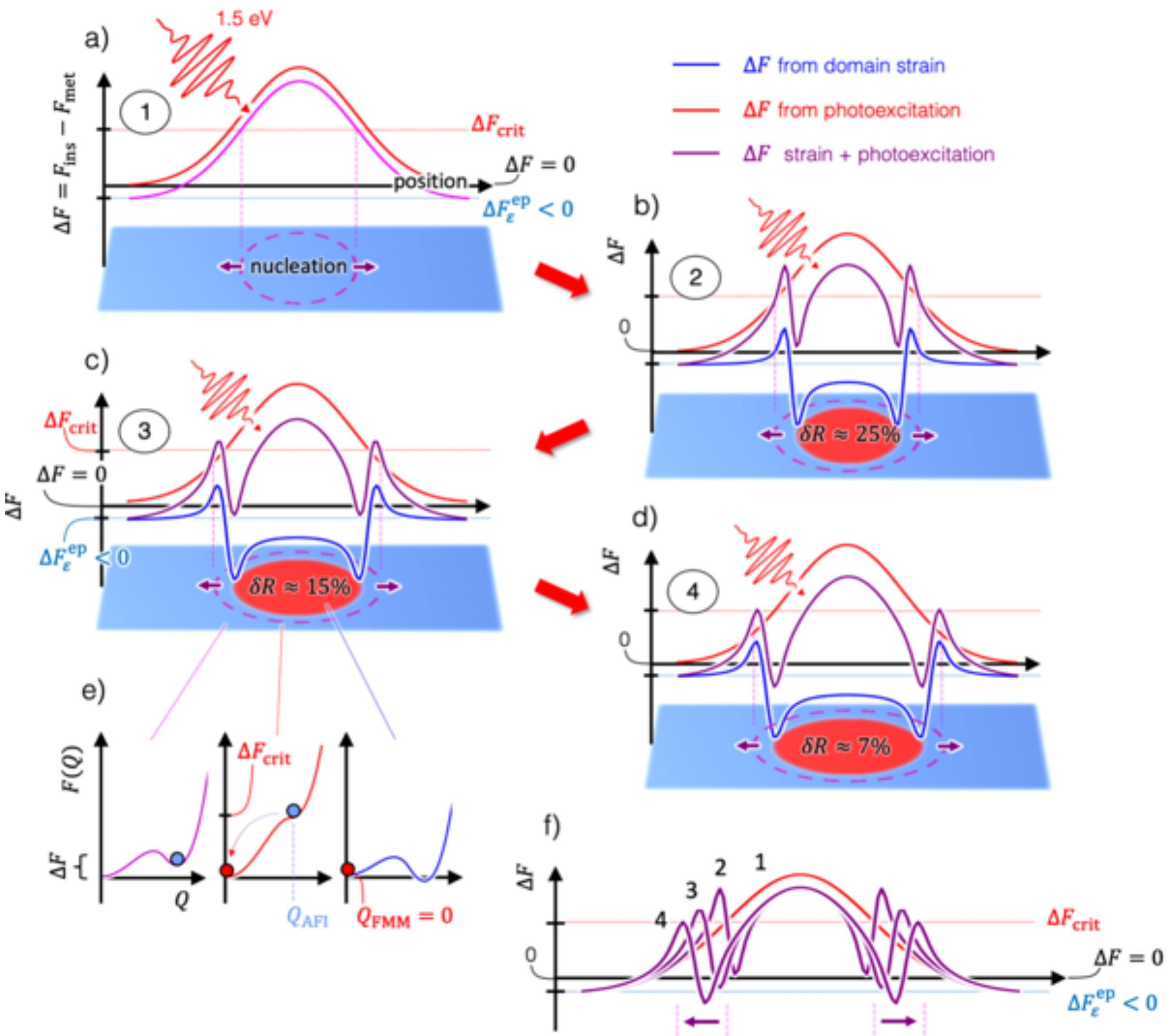

## 8.6 *Figure S6 | Strain-assisted cumulative growth of photo-excited domains.*

**a-d)** Graphics labeled 1-4 sequentially indicate the difference in Landau free energy ($\Delta F$) between antiferromagnetic insulator (AFI) and ferromagnetic metal (FMM) phases as a function of position relative to the spatial center of a gaussian-distributed photo-excitation field (red) incident upon the LCMO film. Numbers 1-4 indicate the cumulative number of photoexcitation pulses delivered to the sample. In this simple model, the total energy difference (magenta) results from a sum of the epitaxial strain energy, photoexcitation energy (red), and local strain fields (blue) arising from a pre-existing FMM domain inclusion. The kinetic barrier for photoinduced transition to the FMM phase is removed wherever $\Delta F$ exceeds a critical value $\Delta F_{\text{crit}}$, which occurs for each delivered photoexcitation pulse immediately outside the boundary of the pre-existing FMM domain. This provides a mechanism for strain-assisted growth of the FMM domain, which saturates at a domain radius where the combined effect of domain-induced strain and the gaussian excitation field fail to exceed $\Delta F_{\text{crit}}$. **e)** Example curves of free energy $F$ versus structural distortion $Q$ at locations: i) far outside the FMM domain (magenta) where the kinetic barrier maintains the AFI phase (blue dot); ii) immediately outside the FMM domain where the kinetic barrier is removed by photoexcitation, thus activating the photoinduced insulator-metal transition (blue dot to red dot); and iii) within the FMM domain where the kinetic barrier maintains the metastable FMM phase (red dot). **f)** Comparative presentation of total $\Delta F$ versus position for each of the scenarios 1-4, showing sequential enlargement of the region for which $\Delta F$ exceeds the critical value $\Delta F_{\mathrm{crit}}$, inducing cumulative growth of the FMM domain.

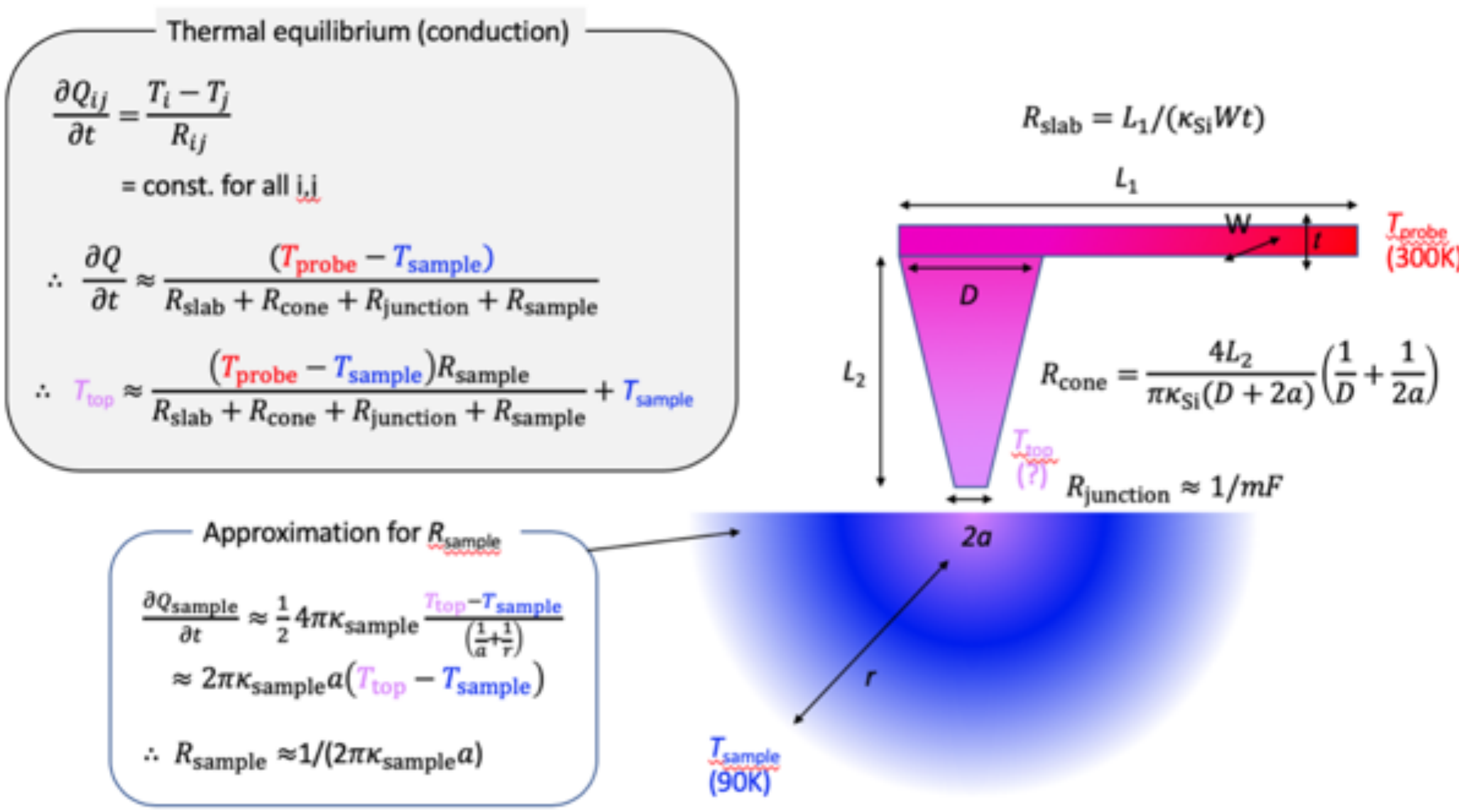

## 8.7 *Figure S7 | Simple model of probe-sample thermal conductance to the LCMO film.*

**(left, top)** A simple model of thermal conductance between the AFM probe and the LCMO film is given in terms of thermal resistances for the constituent elements when the probe is in physical contact with the sample over a contact area of radius $a$ established from Hertzian mechanics. The relevant temperature for establishing typical heating of the LCMO film is $T_{top}$ at the contact point with the AFM probe. **(right)** We consider that the probe holder is at room temperature, and the probe itself comprises a slab-like cantilever and a cone-like tip shaft terminating at a thermally resistive junction defined by rough atomistic contact with the sample. The sample temperature at a quasi-infinite distance is that registered by the temperature probe, 90K. **(bottom, left)** A simple model for thermal resistance of the sample under the assumption that its thermal conductivity is homogeneous.

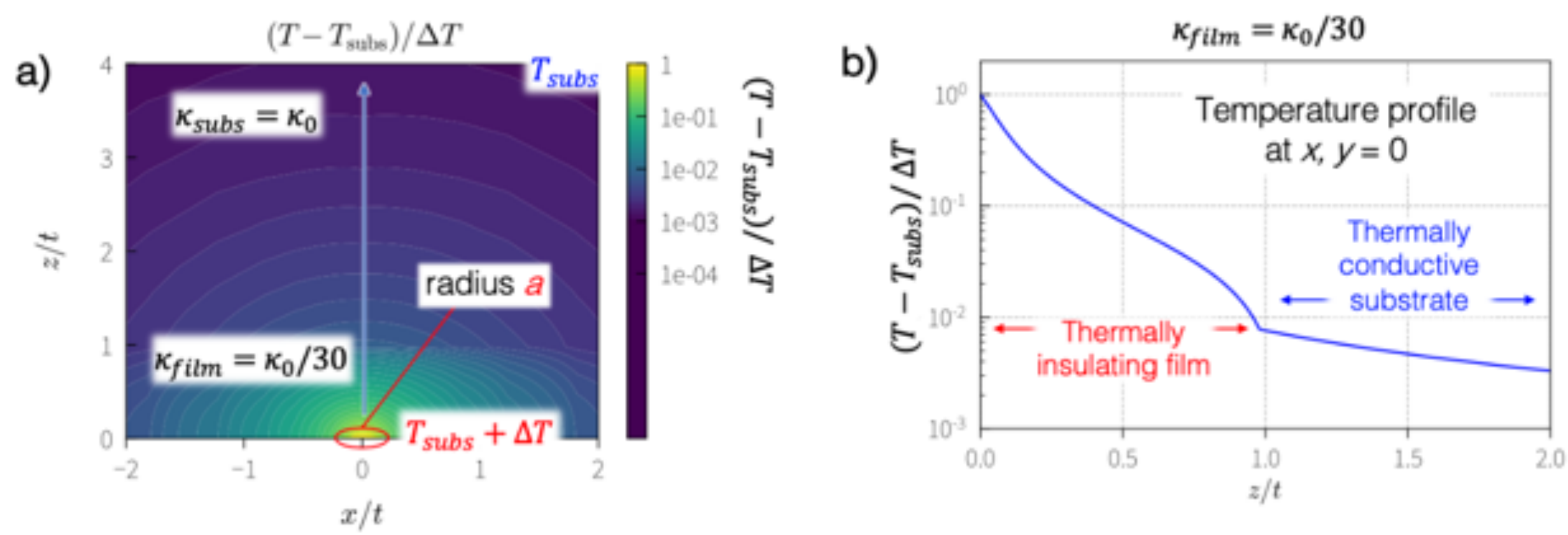

## 8.8 *Figure S8 | Realistic temperature distribution within the LCMO film.*

**a)** A cross-sectional view at $y = 0$ of a quasi-infinite 3-dimensional finite element simulation of heat propagation in the LCMO film / NGO substrate heterostructure. This simulation solves the inhomogeneous thermal problem defined by the steady-state heat equation $\nabla \cdot (\kappa \nabla T) = 0$, with $\kappa$ discontinuous between the film ($z \in [0, t]$) and the substrate ($z \in [t, \infty]$), and under the thermal boundary conditions displayed by the red and blue temperatures. Taking $\Delta T = T_{\text{top}} - T_{\text{subs}}$, this thermal simulation is used to estimate the thermal resistance of the sample heterostructure for the specific choice of a probe-sample contact radius $a$. **b)** A line profile at $x, y = 0$ demonstrating that temperature drop is predominantly within the thermally resistive LCMO film.

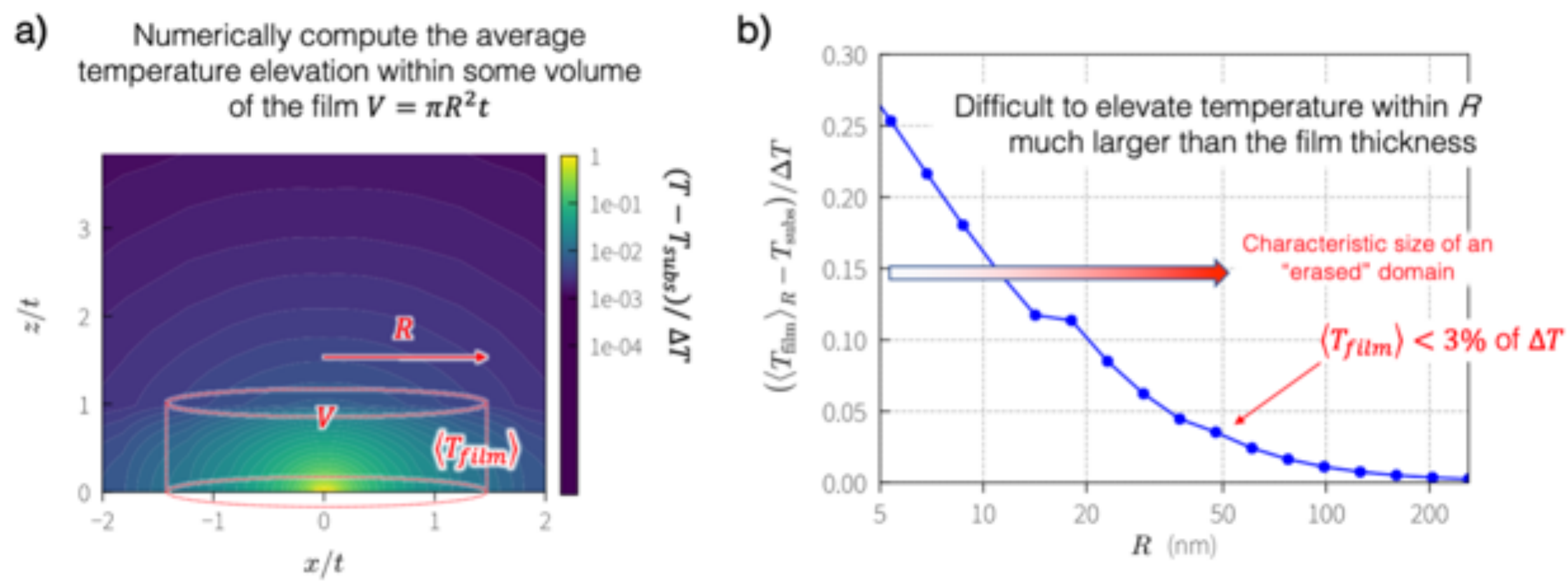

## 8.9 *Figure S9 | Temperature increase within a finite volume of the LCMO film.*

**a)** The temperature distribution for the film / substrate heterostructure as solved by a finite element simulation of heat transport, indicating schematically the cylindrical volume in which the average temperature of the film should be considered. **b)** Average temperature increases in the film averaged over increasingly large cylindrical radii, relative to the temperature difference $\Delta T = T_{\text{top}} - T_{\text{subs}}$ between the probe-sample contact point and the substrate temperature at quasi-infinite distance. At relevant length scales such as $R \approx 50$nm the expected temperature increase is several percent of $\Delta T$.

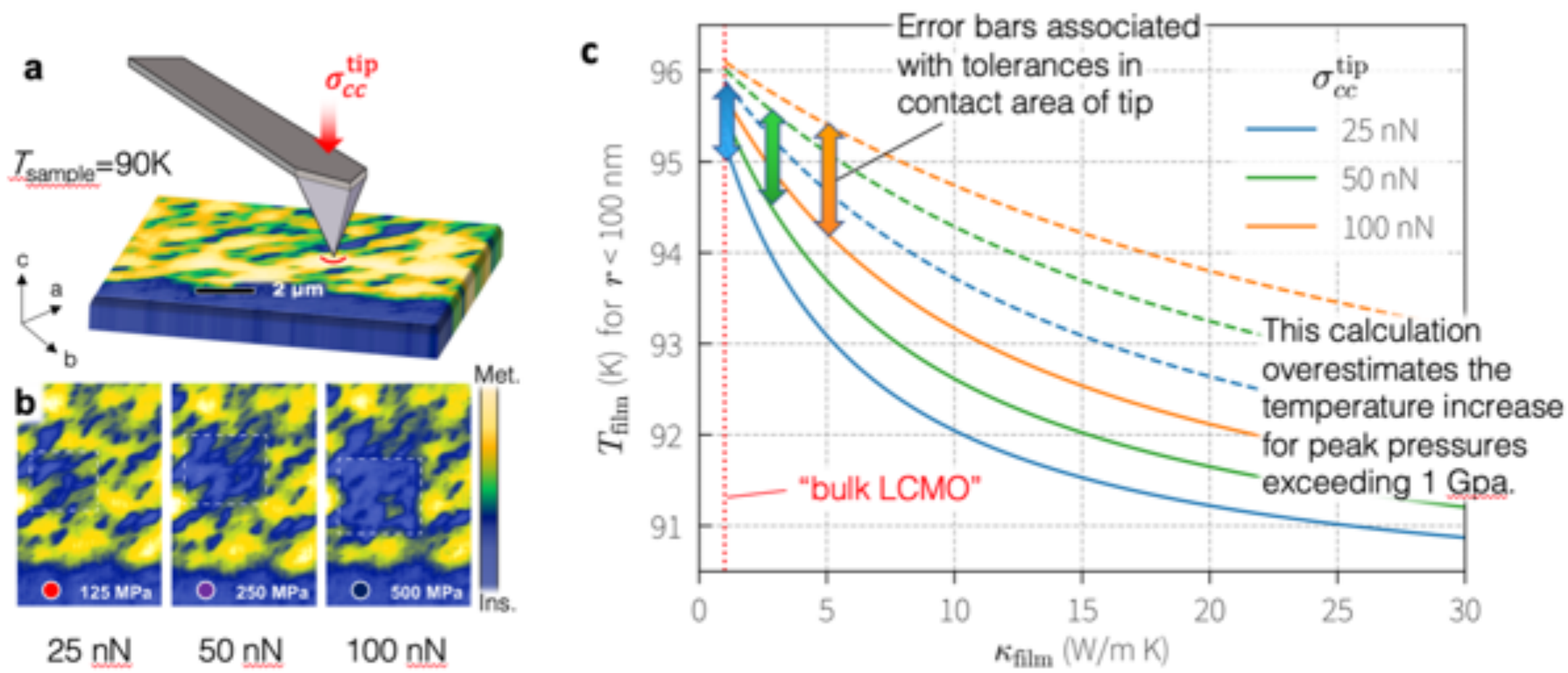

## 8.10 Figure S10 | Maximum heating within a 100 nm metallic domain.

**a, b)** Reproduction of images in Fig. 5 of the main text displaying "erasure" of the photoinduced ferromagnetic metal phase at increasing probe-sample contact forces. **c)** Synthesizing elements of the model for thermal conduction between the AFM probe and the sample, we estimate temperature increases within relevant volumes of the LCMO film for characteristic probe-sample forces used in the experiment. For thermal conductivities of the sample as low as that typical of "bulk" LCMO, the typical temperature increase is no more than 6 degrees K, less than necessary to reach the melting temperature $T_M \approx 105\text{K}$. These estimations overestimate thermal conductance at the probe-sample contact junction, and also disregard the magnetic metallic film deposited over the AFM probe which is should to further reduce thermal conductance at the junction. We therefore regard these estimations as a "best case" for local heating of the LCMO film.

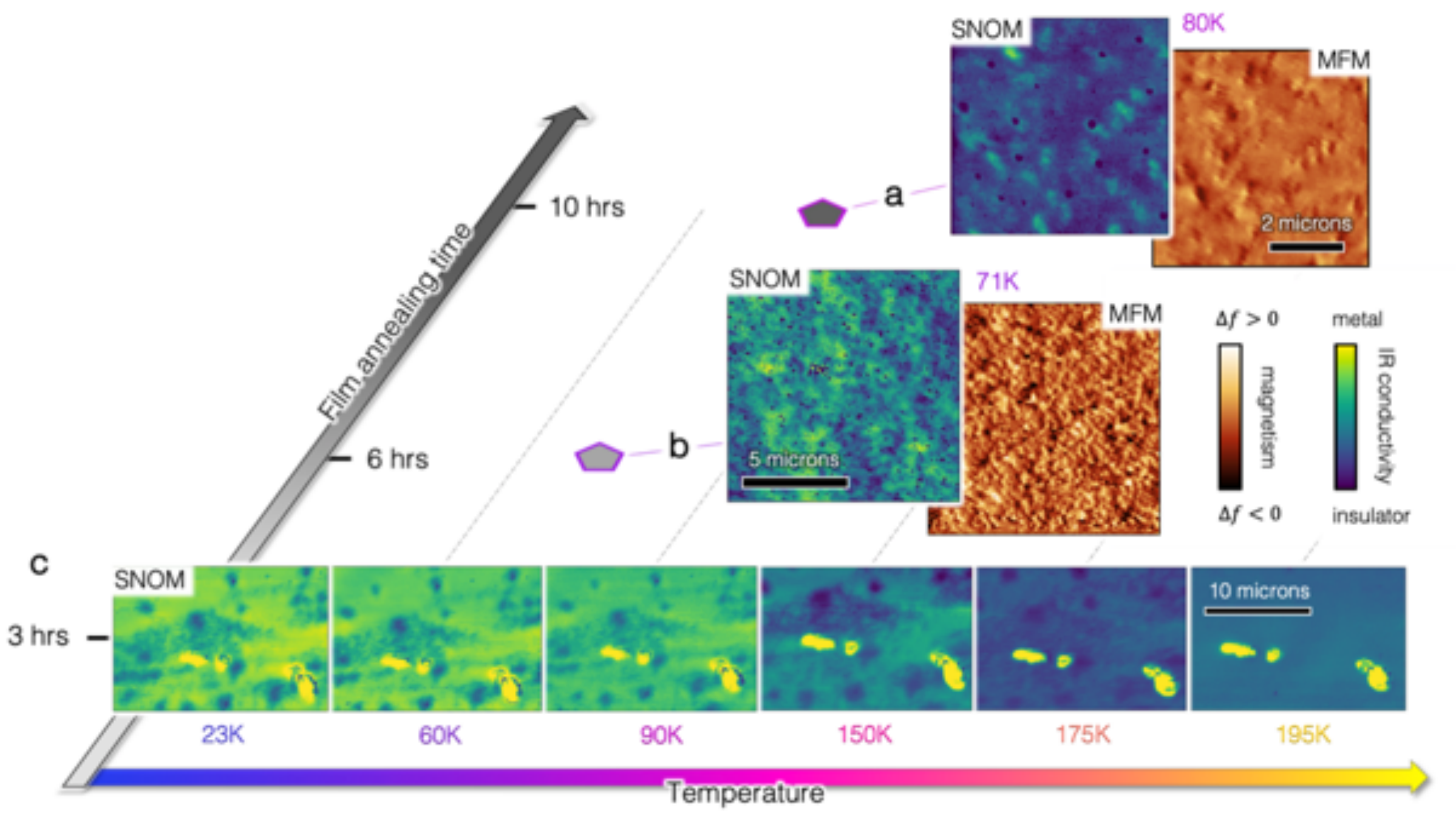

## 8.11 Figure S11 | *Infrared nano-imaging and magnetic force microscopy of LCMO films at varied coherence of epitaxial strain.*

**a)** Scanning near-field optical microscopy (SNOM) and magnetic force microscopy (MFM) images for the LCMO film discussed in the main text, as acquired in a 5x5 micron area at $T$=80K, showing sparse "islands" of ferromagnetic metal in an insulating background as typical for films annealed in oxygenated atmosphere for 10 hrs or more. **b)** SNOM and MFM images for an otherwise identical LCMO film annealed at 6 hrs, acquired in a 10x10 micron area at $T$=71K. Ferromagnetic metal is found to emerge in coexistence with a (putatively) antiferromagnetic insulating phase, both at roughly equal phase fractions, for temperatures below approximately 100K. **c)** SNOM images of an LCMO film annealed for only 3 hrs characterized by comparatively incoherent epitaxial strain from the NGO substrate. Three "islands" of exceptionally high infrared (IR) conductivity (appearing bright yellow) are regions of gold sputtered onto the film surface and used as an optical reference to establish a fixed color scale for IR conductivities across the present images. Warming the sample from $T$=23K displays "melting" of the nearly homogeneous ferromagnetic (presumably) metal phase at a Curie temperature of about 180K, in accord with resistive transport measurements of the same film (not shown). These data inform the phase diagram presented in Fig. 5c of the main text.